\documentclass[11pt]{article}

\usepackage{fullpage}
\usepackage{color}
\usepackage{amssymb,xspace,latexsym,epic,eepic,graphicx}
\usepackage{epsfig,amsmath}
\usepackage{hhline}
\usepackage{times}
\usepackage[lflt]{floatflt}
\usepackage{multicol}
\usepackage[lflt]{floatflt}
\usepackage{setspace}
\usepackage{threeparttable}

\newtheorem{example}{Example}
\newtheorem{definition}{Definition}
\newtheorem{lemma}{Lemma}
\newtheorem{remark}{Remark}
\newtheorem{theorem}{Theorem}
\newtheorem{proposition}{Proposition}
\newtheorem{corollary}{Corollary}

\newcommand{\hproof}[1]{{\noindent\bf Proof:~~ }#1 \boxtheorem
\newline \newline
\noindent} 

\newcommand{\boxtheorem}{\hfill $\Box$}
\newcommand{\ignore}[1]{}
\newcommand{\nit}[1]{{\it #1}}

\newcommand{\go}{g_\oslash}
\newcommand{\D}{\nit{D}}
\newcommand{\SD}{\nit{D}}

\title{\bf Consistent Query Answering under Spatial Semantic Constraints}

\author{
M. Andrea Rodr\'\i guez \\
Universidad de Concepci\'on, Chile\\
\begin{normalsize}
\texttt{andrea@udec.cl}
\end{normalsize}
\and
Leopoldo Bertossi\thanks{Faculty Fellow of the IBM Center for Advanced Studies. Also affiliated to Universidad de Concepci\'on, Chile.}\\
Carleton University, Canada\\
\begin{normalsize}
\texttt{bertossi@scs.carleton.ca}
\end{normalsize}
\and
M\'onica Caniup\'an\\
Universidad del B\'\i o-B\'\i o, Chile\\
\begin{normalsize}
 \texttt{mcaniupa@ubiobio.cl}
 \end{normalsize}
}
\date{}

\begin{document}

\maketitle

\begin{abstract}
\noindent Consistent query answering is an inconsistency
tolerant approach to obtaining semantically correct answers
from a database that may be inconsistent with respect to its
integrity constraints. In this work we formalize the notion of
consistent query answer for spatial databases and spatial
semantic integrity constraints. In order to do this, we first
characterize conflicting spatial data, and next, we define
admissible instances that restore consistency while staying
close to the original instance. In this way we obtain a repair
semantics, which is used as an instrumental concept to define
and possibly derive consistent query answers. We then concentrate on a class of spatial denial constraints and spatial queries for which there exists an efficient strategy to compute consistent query answers. This study applies inconsistency tolerance in spatial databases, rising research issues that shift the goal from the consistency of a spatial database to the consistency of query answering.
\end{abstract}

\section{Introduction}\label{se:introduction}

Consistency in database systems is defined as the satisfaction
by a database instance of a set of integrity constraints (ICs)
that restricts the admissible database states. Although consistency
is a desirable and usually enforced property of databa\-ses, it
is not uncommon to find inconsistent spatial databases  due to
data integration, unforced integrity constraints, legacy data,
or time lag updates.  In the presence  of inconsistencies,
there are alternative courses of action: (a) ignore
inconsistencies, (b) restore consistency via updates on the
database, or (c) accept inconsistencies, without changing the
database, but compute the ``consistent or correct" answers to
queries.  For many reasons, the first two alternatives may not
be appropriate \cite{BC03}, specially in the case of virtual
data integration \cite{BB04IT}, where
 centralized and global changes to the data sources are not allowed.
The latter alternative has been investigated in the relational
case  \cite{bertossiSigRec06,chomickiICDT07}. In this paper we
explore this approach in the spatial domain, i.e., for spatial
databases and with respect to spatial semantic integrity
constraints (SICs).

Extracting  consistent data from inconsistent databases  could
be qualified as an ``inconsistency tolerant" approach to
querying databases. A piece of data will be part
of a consistent answer if it is not logically related to the
inconsistencies in the database with respect to its set of ICs.
We introduce this idea using an informal and simple example.

\begin{example}  \label{ex:motivating} \em
Consider a database instance with a relation \textit{LandP},
denoting land parcels, with a thematic attribute (${\nit
idl}$), and a spatial attribute, ${\nit geometry}$, of data
type ${\nit polygon}$. An IC stating that geometries of two
different land parcels must be disjoint or just touch, i.e.,
they cannot internally intersect, is expected to be satisfied.
However, the instance in Figure \ref{fig:motiving} does not
satisfy this IC and therefore it is inconsistent: ~the land
parcels with idls ${\it idl}_2$ and ${\it idl}_3$ overlap.
Notice that these geometries are partially in conflict and what
is not in conflict can be considered as consistent data.

\begin{figure}[ht!]
\begin{center}
\begin{multicols}{2}
\begin{small}
\begin {tabular} {|cc|}
\hline\multicolumn{2}{|c|}{\textbf{LandP}}\\\hline
\textit{idl}&{geometry}\\\hline
$idl_1$&$g_1$\\
$idl_2$&$g_2$\\
$idl_3$&$g_3$\\
\cline{1-2}
\end {tabular}\\
\end{small}
~~~\begin{tabular}{c}
\includegraphics[width=2.5 cm]{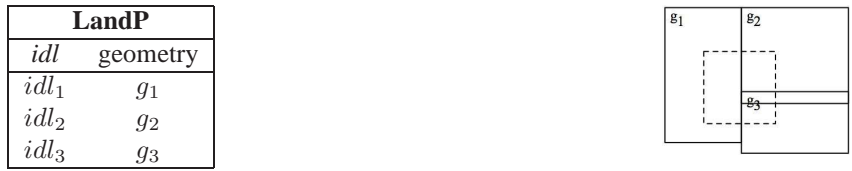}\\
\end{tabular}\\
\end{multicols}
\vspace{-4mm}\caption{~An inconsistent spatial database}
\label{fig:motiving}
\end{center}
\end{figure}

\noindent Suppose that a query requests all land parcels whose
geometries intersect with a query window, which represents the
spatial region shown in Figure \ref{fig:motiving} as a
rectangle with dashed borders. Although the database instance
is inconsistent, we can still obtain useful and meaningful
answers. In this case, only the intersection of $g_2$ and $g_3$
is in conflict, but the rest of both geometries can be
considered consistent and should be part of any ``database
repair" if we decide to  restore consistency by means of
minimal geometric changes. Thus, since the non-conflicting
parts of geometries $g_2$ and $g_3$ intersect the query window,
we would expect an answer including land parcels with
identities ${\it idl_1,idl_2}$ and ${\it idl_3}$. \boxtheorem
\end{example}

\noindent If we just concentrate on (in)consistency issues in
databases (leaving aside consistent query answering for a
moment), we can see that, in contrast to (in)consistency
handling in relational databases, that has been largely
investigated, not much research of this kind has been done for
spatial databases. In particular, there is not much work around
the formalization of semantic spatial ICs, satisfaction of ICs, and checking and
maintenance of ICs in the spatial domain. However, some papers
address the specification of some kinds of integrity
constraints \cite{BORLAE99,HADTRY92}, and checking topological
consistency at multiple representations and for data
integration \cite{EGECLE94,EGESHA93,TRYEGE1997}.

More recently, \cite{DBLP:journals/isci/DuckhamLMW06} proposes
qualitative reasoning with description logic to describe
consistency between geographic data sets.  In \cite{MAS07}  a
set of abstract relations between entity classes is defined;
and they could be used  to discover redundancies and conflicts
in sets of SICs.  A  proposal for fixing (changing) spatial
database instances under different types of spatial
inconsistencies is given in \cite{SERUB00}. According to it,
changes are applied over geometries in isolation; that is, they
are not analyzed in combination with multiple  SICs. In
\cite{RODRI05} some issues around query answering under
violations of functional dependencies involving geometric
attributes were raised. However, the problem of dealing with an
inconsistent spatial database, while still obtaining meaningful
answers, has not been systematically studied so far.

Consistent query answering (CQA) from inconsistent databases as a strategy of inconsistent tolerance has an extensive literature (cf.
\cite{bertossiSigRec06,BC03,chomickiICDT07} for surveys). It
was introduced and studied in the context of relational
database in \cite{ABC99}. They defined consistent answers to queries as those that are invariant under all the minimal forms of restoring consistency of the original database. Thus, the notion of {\em repair} of
an instance with respect to a set of ICs becomes a fundamental
concept for defining consistent query answers. A {\it repair
semantics} defines the admissible and consistent alternative
instances to an inconsistent database at hand. More precisely,
a {\em repair} of an inconsistent relational  instance $\D$ is
a consistent instance $\D'$ obtained from $\D$ by deleting or
inserting whole tuples. The set of tuples by which $\D$ and
$\D'$ differ is minimal under set inclusion \cite{ABC99}. Other
types of repair semantics have been studied in the relational
case. For example, in
\cite{FLLPS01,DBLP:journals/tods/Wijsen05} repairs are obtained
by allowing updates of attribute values in tuples.

In comparison to the relational case, spatial databases  offer new
alternatives and challenges when defining a repair semantics. This
is due, in particular, to the use of complex attributes to
represent geometries, their combination with thematic attributes,
and the nature of spatial (topological) relations.

In this work we define a repair semantics for spatial databases
with respect to a subset of spatial semantic integrity constraints (a.k.a.
topo-semantic integrity constraints) \cite{SERUB00}, which
impose semantic restrictions on topological predicates and combinations thereof. In particular, we treat spatial semantic integrity constraints that can be expressed by denials constraints.
For example, they can specify that ``two land parcels cannot internally intersect". This class
of constraints are neither standardized nor
integrated into current spatial database management systems
(DBMSs); they rather depend on the application, and must be defined and handled by
the database developers. They are very important because they capture the semantics of the intended models.  Spatial semantic integrity constraints  will be simply called
{\em spatial integrity constraints} (SICs). Other spatial
integrity constraints \cite{COCK97} are \textit{domain
(topological or geometric) constraints}, and
 they refer to the geometry, topology, and spatial
relations of the spatial data types. One of them could specify
that ``polygons must be closed''. Many of these geometric constraints are
now commonly integrated into spatial DBMSs \cite{OGIS99}.

A definition of a repair semantics for spatial DBs and CQA for
spatial range queries was first proposed in \cite{RodBerCani08}, where we discussed the idea of shrinking geometries to solve conflicting tuples and applied to CQA for range queries. In this paper we complement and extend
our previous work with the following  main contributions: (1)
We  formalize the repair semantics of a spatial database
instance under violations of SICs. This is done  through
virtual changes of geometries that participate in violations of
SICs. Unlike \cite{RodBerCani08}, we identify the admissible
local transformations and we use them to provide an inductive
definition of database repair. (2) Based on this formalization,
a consistent answer to a spatial query is defined as an answer
obtained from all the admissible repairs. Extending the results
in \cite{RodBerCani08}, we now define CQA not only for range
but also for spatial join queries. (3) Although the repair
semantics and consistent query answers can be defined for a
fairly broad class of SICs and queries, as it becomes clear
soon, naive algorithms for computing consistent answers on the
basis of the computation of all repairs are of exponential
time. For this reason, CQA for a relevant subset of SICs and
range and join  queries is done via a {\em core computation}.
This amounts to querying directly the  intersection of all repairs
of an inconsistent database instance, but without actually
computing the repairs. We show cases where this core can be
specified as a view of the original, inconsistent database.~(4)~
We present an experimental evaluation with real and synthetic data sets that compares the cost of
CQA with the cost of evaluating queries directly over the
inconsistent database (i.e., ignoring inconsistencies).

The rest of the paper is organized as follows. In Section
\ref{se:preliminaries} we describe the spatial data model  upon
which we define the repair semantics and consistent query
answers. A formal definition of repair for spatial inconsistent
databases under SICs is introduced in Section~\ref{se:repairs}.
In Section~\ref{se:CQA} we define consistent answers to
conjunctive queries. We analyze  in particular the cases of
range and join queries with respect to their computational properties. This
leads us, in Section~\ref{sec:core}, to propose polynomial time algorithms (in data
complexity) for consistent query answering with respect to a relevant class
of SICs and queries. An experimental
evaluation of the cost of CQA is provided in
Section~\ref{sec:experiments}. Final conclusions and future
research directions are given in Section~\ref{se:conclusions}.

\section{Preliminaries}\label{se:preliminaries}

Current models of spatial database are typically seen as extensions
of the relational data model (known as extended-relational or object-relational models)
with the definition of abstract data types to specify spatial
attributes. We now introduce a general
spatio-relational database model that includes
spatio-relational predicates (they could also be purely
relational) and spatial ICs. It uses
some of the definitions introduced in \cite{paradaens98data}.
The  model is independent of the geometric data model (e.g.
Spaghetti \cite{THOMLAURI92}, topological
\cite{DBLP:conf/vldb/Guting94,THOMLAURI92}, raster
\cite{DBLP:journals/vldb/GutingS95}, or polynomial model
\cite{pare94}) underlying the
representation of spatial data types.

A {\em spatio-relational database schema} is of the form
$\Sigma=({\cal U}, {\cal A}, {\cal R},$  ${\cal T}, {\cal O},
{\cal B})$, where: (a) ${\cal U}$ is the possibly infinite
database domain of atomic thematic values. (b) ${\cal A}$ is a
set of thematic, non-spatial, attributes. (c) ${\cal R}$ is a
finite set of spatio-relational predicates whose attributes
belong to ${\cal A}$ or are spatial attributes. Spatial attributes take
admissible values in ${\cal P}(\mathbb{R}^m)$, the power set of
$\mathbb{R}^m$, for an $m$ that depends on the dimension of the spatial attribute. (d)
${\cal T}$ is a fixed set of binary spatial predicates, with a built-in interpretation. (e)
${\cal O}$ is a fixed set of geometric operators that take
spatial arguments, also with a built-in interpretation. (f) ${\cal B}$ is a fixed set of built-in
relational predicates, like comparison predicates, e.g.
$<,>,=,\neq$, which apply to thematic attribute values.

Each database predicate $R \in {\cal R}$ has a type $\tau(R) =
[n,m]$, with $n,m \in \mathbb{N}$, indicating the number $n$ of
thematic attributes, and the spatial dimension $m$ of the
single spatial attribute (it takes values in ${\cal
P}(\mathbb{R}^m$)).\footnote{For simplicity, we use one spatial
attribute, but it is not difficult to consider a greater number
of spatial attributes.} In Example \ref{ex:motivating},
$\tau({\it LandP}) = [1,2]$, since it has one thematic
attribute (${\it idl}$) and one spatial attribute (${\it
geometry}$) defined by a 2D polygon. In this work we assume
that each relation $R$ has a  key of the form
(\ref{eq:primarykey}) formed by thematic attributes only:
\begin{equation}\label{eq:primarykey}
 \forall \bar{x}_1 \bar{x}_2 \bar{x}_3 s_1 s_2 ~(
R(\bar{x}_1,\bar{x}_2;s_1) \wedge R(\bar{x}_1,\bar{x}_3;s_2)  \rightarrow (\bar{x}_2 = \bar{x}_3 \wedge s_1=s_2)),
\end{equation}
where the $\bar{x}_i$ are sequences of distinct variables
representing thematic attributes of $R$, and the $s_i$ are
variables for geometric attributes. Here $s_1 = s_2$ means geometric equality; that is, the identity of two geometries.

A database instance $\D$ of a spatio-relational schema $\Sigma$
is a finite collection of ground atoms (or {\em spatial
database tuples}) of the form $R(c_{1},...,c_{n};s)$, where $R
\in {\cal R}$, $\langle c_{1},...,c_{n}\rangle \in {\cal U}^n$
contains the thematic attribute values, and  ${\it s} \in
\nit{Ad} \subseteq {\cal P}({\mathbb R}^m)$, where $\nit{Ad}$
is the class of admissible geometries (cf. below). The extension
in a particular instance of a spatio-relational predicate  is a
subset of ${\cal U}^n \times \nit{Ad}$. For simplicity, and to
fix ideas, we will consider the case where  $m = 2$.

Among the different abstraction mechanisms  for modelling single
spatial objects, we concentrate on  regions for modelling real objects that
have an extent. They are useful in a broad class of applications
in Geographic Information Systems (GISs). More specifically,  our
model will be compatible with the specification of spatial operators (i.e., spatial relations or geometric operations) as found in current spatial DBMSs \cite{OGIS99}. Following current implementations of DBMSs, regions could be defined as finite sets of polygons that,
in their turn, are defined through their vertices. This would make regions finitely representable. However, in this work  geometries will be treated at a more abstract level, which is independent of the spatial model  used for geometric representation. In consequence, an admissible geometry of the Euclidean plane is either the  empty geometry, $\go$,
which corresponds to the empty subset of the plane, or is a
closed and bounded region with a positive area. It holds $\go \cap g = g \cap \go = \go$, for every region $g$. From now on, empty geometries and regions of $\mathbb{R}^2$ are called {\em
admissible geometries} and they form the class $\nit{Ad}$.

Geometric attributes are complex data types, and their
manipulation may have an important effect on the computational
cost of certain algorithms and algorithmic problems. As usual,
we are interested in {\em data complexity}, i.e., in terms of
the size of the database. The {\em size} of a spatio-relational
database can be  defined  as a function of the number of tuples and
the representation size of geometries in those tuples.

\begin{figure}[t!]
\begin{center}
\includegraphics[width=10cm]{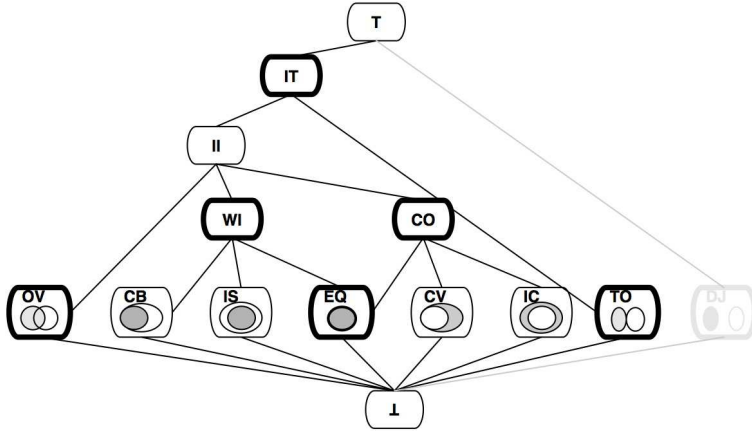}
\caption{~Subsumption lattice of topological relations between regions: OV
(Overlaps), CB (CoveredBy),  IS (Inside), EQ (Equals), CV
(Covers), IC (Includes), TO (Touches), DJ (Disjoint), IT
(Intersects), II (IIntersects), WI (Within), and CO (Contains).}
\label{lattice}
\end{center}
\end{figure}

We concentrate on  binary (i.e., two-ary) spatial predicates
that represent topological relations between {\it regions}. They have a fixed semantics, and become the elements
of ${\cal T}$. There are eight basic binary relations over regions of $\mathbb{R}^2$: ${\it Overlaps~(OV)}$,
${\it Equals~(EQ)}$, ${\it CoveredBy~(CB)}$, ${\it
Inside~(IS)}$, ${\it Covers}$ ${\it (CV)}$,  ${\it
Includes~(IC)}$, ${\it Touches~(TO)}$, and ${\it
Disjoint~(DJ)}$
\cite{EGENFRAN91,RCC92}.\footnote{The names of relations chosen
here are in agreement with the names used in current SQL
languages \cite{OGIS99}, but differ slightly from the names
found in the research literature. The relations found in
current SQL languages are represented in Figure \ref{lattice}
with thick borders.} The semantics of the  topological relations follows the point-set topology defined in \cite{EGENFRAN91}, which is not defined for empty geometries. We will apply this semantics to our non-empty admissible
geometries. For the case of the empty set, a separate definition will be given below. According to \cite{EGENFRAN91}, an atom $T(x,y)$
becomes true if four conditions are simultaneously true. Those conditions are expressed in terms of
emptyness ($\emptyset$) and non-emptyness ($\neg \emptyset$) of the intersection of their boundaries ($\delta$) and interiors ($\circ$). The definitions
can be found in  Table~\ref{tab:relations}. For example, for non-empty regions $x,y$, $\nit{TO}(x,y)$ is true iff
all of $\delta(x) \cap \delta(y) \neq \emptyset$, $\circ(x) \cap \circ(y) = \emptyset$, $\delta(x) \cap \circ(y) = \emptyset$, and $\circ(x) \cap \delta(y) = \emptyset$ simultaneously hold.

\begin{table}[h]
\begin{center}
\begin{tabular}{lcccc}
\hline
Relation&$\delta(x) \cap \delta(y)$&$\circ(x) \cap \circ(y)$&$\delta(x) \cap \circ(y)$&$\circ(x) \cap \delta(y)$\\
\hline
DJ(x,y)&$\emptyset$&$\emptyset$&$\emptyset$&$\emptyset$\\
TO(x,y)&$\neg \emptyset$&$\emptyset$&$\emptyset$&$\emptyset$\\
EQ(x,y)&$\neg \emptyset$&$\neg \emptyset$&$\emptyset$&$\emptyset$\\
IS(x,y)&$\emptyset$&$\neg \emptyset$&$\neg \emptyset$&$\emptyset$\\
CB(x,y)&$\neg \emptyset$&$\neg \emptyset$&$\neg \emptyset$&$\emptyset$\\
IC(x,y)&$\emptyset$&$\neg \emptyset$&$\emptyset$&$\neg \emptyset$\\
CV(x,y)&$\neg \emptyset$&$\neg \emptyset$&$\emptyset$&$\neg \emptyset$\\
OV(x,y)&$\neg \emptyset$&$\neg \emptyset$&$\neg \emptyset$&$\neg \emptyset$\\
\end{tabular}
\caption{Definition of topological relations between regions based on point-set topology} \label{tab:relations}
\end{center}
\end{table}

\noindent In this work we exclude the topological relation ${\nit Disjoint}$ from ${\cal T}$. This decision is discussed in  Section~\ref{se:repairs}, where we introduce the repair semantics. In addition to the basic topological relations, we consider
three {\em derived relations} that exist in current SQL
languages and can be logically defined in terms of the other
basic predicates: ${\it Intersects~(IT)}$, ${\it Within~(WI)}$,
and ${\it Contains~(CO)}$. We also introduce a forth relation,
{\it IIntersects~(II)}, that holds when the interiors of two
geometries intersect. It can be logically defined  as the
disjunction of  ${\it Overlaps}$, ${\it Within}$ and ${\it
Contains}$  (cf. Figure \ref{lattice}). For all the topological relations in ${\cal T}$, their
converse (inverse) relation is within the set. Some of them are
symmetric,  like  ${\it Equals}$, ${\it
Touches}$, and ${\it Overlaps}$. For the non-symmetric
relations,  the converse relation of ${\it CoveredBy}$ is ${\it
Covers}$, of ${\it Inside}$ is ${\it Includes}$, and of  ${\it Within}$ is ${\it Contains}$.

As mentioned before, the formal definitions of the topological relations \cite{EGENFRAN91,RCC92} do not consider the empty geometry as an argument. Indeed, at the best of our knowledge, no clear semantics for topological predicates with empty geometries exists. However, in our case we extent the definitions in order to deal with this
 case. This will allow us to use a classical bi-valued logic, where atoms are always true or false, but never undefined. According to our extended definition,  for any $T \in {\cal T}$, $T(g_1,g_2)$ is false if $g_1 = \go$ or $g_2 =\go$.
 In particular, $\nit{IS}(g,\go)$ is false, for every admissible region $g$. In order to make comparisons with the empty region, we will introduce and  use a special predicate $\nit{IsEmpty}(\cdot)$ on admissible geometries, such that $\nit{IsEmpty}(s)$ is true iff $s = \go$.

Notice that the semantics of the topological predicates, even for non-empty regions, may differ
from the  intuitive set-theoretic semantics one could assign to them. For example, for an admissible
and non-empty geometry $g$, $\nit{OV}(g,g)$ is false (due to the conditions in the last two columns in Table
 \ref{tab:relations}). In consequence, the constraint $\forall x \forall s^{\neq \go} \neg (R(x;s) \wedge \nit{OV}(s,s))$ is satisfied.

Given a database instance, additional spatial information is
usually computed from the explicit geometric data by means of
the spatial operators in ${\cal O}$ associated with $\Sigma$.
Some relevant operators are: ${\it Area}$, ${\it Union}$
(binary), ${\it Intersection}$, ${\it Difference}$, ${\it
Buffer}$, and ${\it Union}$ ${\it Aggregation}$ ${\it
(GeomUnion)}$.\footnote{Operator ${\it GeomUnion}$ returns the
geometry that represents the point set union of all geometries
in a given set, an operator also known as a spatial aggregation operator. Although this function is part of SQL for
several spatial databases (Postgres/PostGIS, Oracle), it is not
explicitly defined in the OGC specification~ \cite{OGIS99}.}
(Cf. \cite{OGIS99} for the complete set of spatial predicates
defined within  the Open GIS Consortium.) There are several spatial operators used in this work; however, we will
identify a particular subset ${\cal O}^a$ of spatial operators
in ${\cal O}$, i.e., ${\cal O}^a \subseteq {\cal O}$, which
will be defined for all admissible geometries and used
to shrink geometries with the purpose of restoring consistency,
as we describe in Section~\ref{se:repairs}.

\begin{definition} \em
\label{de:spatialop} The set ${\cal O}^a$ of {\em admissible
 operations} contains the following geometric
operations on admissible geometries $g$ and
$g'$:\\
\hspace*{.5cm} (1) $\nit{Difference}(g,g')$ is the topological closure of the set-difference.

(2) ${\it Buffer}(g,d)$ is the geometry obtained by
    buffering  a distance $d$ around $g$, where $d$ is a distance unit. ${\it Buffer}(g,d)$ returns a closed region $\bar{g}$ containing geometry $g$, such that every point in the boundary of $\bar{g}$ is
    at a distance $d$ from some point of the boundary of $g$. In particular, ${\it Buffer}(\go,d)=\go$.~
\boxtheorem \end{definition}
Notice that these operators, when applied to
admissible geometries, produce  admissible geometries.

\begin{remark} \label{re:d} \em
The value of $d$  in Definition \ref{de:spatialop} is instance dependent. It  should be precomputed from the spatial input data. For this work, we
consider $d$ to be a fixed value associated with the minimum
distance between geometries in the cartographic scale of the database
instance. \boxtheorem
\end{remark}
A schema $\Sigma$ determines a many-sorted, first-order (FO)
language ${\cal L}(\Sigma)$ of predicate logic. It can be used
to syntactically characterize and express SICs. For simplicity,
we concentrate on {\em denial SICs},\footnote{Denial
constraints are easier to handle in the relational case as
consistency with respect to them is achieved by tuple deletions
only \cite{BC03}.} which are sentences of the form:

\begin{equation}\label{eq:format0}
\forall^{\neq g_\emptyset} \bar{s}\forall \bar{x}~ \neg (\bigwedge_{i = 1}^{m}
R_i(\bar{x}_i;s_i)  \wedge  \varphi  \wedge ~
\bigwedge_{j=1}^nT_j(v_j,w_j)).
\end{equation}
\noindent Here, $\bar{s} = s_1 \cdots s_m,~ \bar{x} = \bar{x}_1
\cdots \bar{x}_m$ are finite sequences of geometric and
thematic variables, respectively, and $0 < m,n \in \mathbb{N}$.
Thus, each $\bar{x}_i$ is a finite tuple of thematic variables and will be treated as a set of attributes, such that $\bar{x}_i \subseteq \bar{x}_j$ means that the variables in $\bar{x}_i$ area also variables in $\bar{x}_j$. Also, $\forall \bar{x}$ stands for $\forall x_1 \cdots \forall x_m$;
and  $\forall^{\neq g_\emptyset} \bar{s}$ stands for $\forall s_1 \cdots \forall s_m$, with the universal quantifiers ranging over all the non-empty admissible geometries (i.e. regions).
Here, $v_j,w_j \in \bar{s}$,  $R_1,\ldots, R_m \in {\cal R}$, $\varphi$
is a formula containing built-in atoms over thematic
attributes, and $T_j \in {\cal T}$. A constraint of the form
(\ref{eq:format0}) prohibits certain combinations of database
atoms. Since topological predicates for empty geometries are always false, the restricted quantification over non-empty geometries in the constraints could be eliminated. However, we do not want to make the satisfaction of the
constraints rely on our particular definition of the topological predicates for the empty region. In this way,
our framework becomes more general, robust and modular, in the sense that it would be possible to redefine the topological
predicates for the empty region without affecting our approach and results.

\begin{figure}[t]
\begin{center}
\begin{multicols}{3}
{\small
\begin{tabular} {|p{0.2cm}p{0.5cm}p{0.6cm}c|} \hline
\multicolumn{4}{|c|}{\textbf{LandP}}\\\hline
{\it idl} & {\it name} & {\it owner} & {\it geometry}\\ \hline
$idl_1$&$n_1$&$o_1$&$g_1$\\
$idl_2$&$n_2$&$o_2$&$g_2$\\
$idl_3$&$n_3$&$o_3$&$g_3$\\
\cline{1-4}
\end{tabular}
}

{\small
\begin{tabular} {|p{0.25cm}c|} \hline
\multicolumn{2}{|c|}{\textbf{Building}}\\ \hline
{\it idb} & {\it geometry}\\
\hline
$idb_1$&$g_4$\\
$idb_2$&$g_5$\\
 & \\
\cline{1-2}
\end{tabular}
}
\includegraphics[width=3cm]{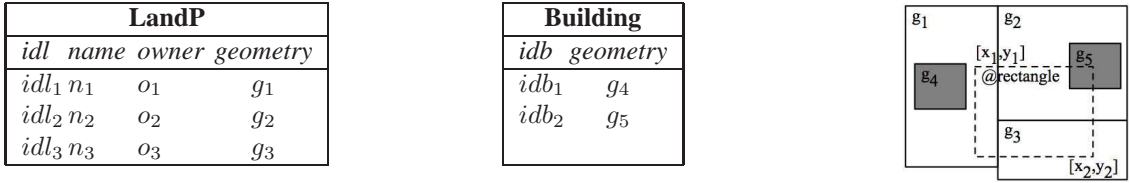}
\end{multicols}
\end{center}
\caption{~A spatial database instance} \label{conEx}
\end{figure}

\begin{example} \em
\label{ex:SDB} Figure \ref{conEx} shows an instance for the
schema ${\cal R} = \{{\it LandP(idl,}$ ${\it name},$ ${\it
owner};$ ${\it geometry)}$, ${\it Building(idb;}$ ${\it
geometry)}\}$. Dark rectangles represent buildings and white
rectangles represents land parcels. In ${\it LandP}$, the
thematic attributes are ${\it idl, name}$ and ${\it owner}$,
whereas ${\it geometry}$ is the spatial attribute of dimension
$2$. Similarly for ${\it Building}$, which has  only ${\it
idl}$ as a thematic attribute.

The following sentences are denial SICs: ~(The symbol $\bar{\forall}$
stands for the universal closure of the formula that follows it.)
\begin{eqnarray}
\overline{\forall}\neg(\nit{LandP}(idl_{1},n_{1},o_{1};s_1) \wedge
\nit{LandP}(idl_{2},n_{2},o_{2};s_2) ~\wedge idl_{1} \neq idl_{2}\wedge  \nit{IIntersects}(s_1,s_2)). \label{eq:constraint}
\end{eqnarray}
\begin{equation}
{\footnotesize\overline{\forall}\neg (\nit{Building}(idb;s_1)~\wedge~ {\it
LandP}(idl,n,o; s_2)~\wedge \nit{Overlaps}(s_1,s_2)).}\label{eq:build}
\end{equation}
The SIC (\ref{eq:constraint}) says that geometries of land
parcels with different {\em ids} cannot internally intersect
(i.e., they can only be disjoint or touch). The SIC
(\ref{eq:build}) establishes that building blocks cannot
(partially) overlap land parcels. \boxtheorem
\end{example}

\noindent A database instance $\D$ for schema $\Sigma$ can be
seen as an interpretation structure for the  language ${\cal
L}(\Sigma)$. For a  set $\Psi$ of SICs in ${\cal L}(\Sigma)$,
~$\D \models \Psi$ denotes  that each of the constraints  in
$\Psi$ is true in (or satisfied by) $\D$. In this case, we say
that $\D$ is {\em consistent} with respect to $\Psi$.
Correspondingly, $\D$ is {\em inconsistent} with respect to $\Psi$, denoted $\D \not \models \Psi$, when there is a $\psi \in \Psi$ that is
{\em violated} by $\D$, i.e., not satisfied by $\D$.  The
instance in Example \ref{ex:SDB} is consistent with respect to
its SICs.

In what follows, we will assume that the set $\Psi$ of SICs under consideration is logically consistent; i.e., that there exists a non-empty database instance $\D$ (not necessarily the one at hand), such that $\D\models \Psi$. For example, any set of SICs containing a constraint of the form $\forall^{\neq g_\emptyset} s\forall \bar{x}~ \neg (R(\bar{x};s)  \wedge {\nit Equals} (s,s))$ is logically inconsistent. The analysis of whether a set of SICs is logically consistent or not is out of the scope of this work.

\section{A Repair Semantics} \label{se:repairs}
Different alternatives for update-based consistency restoration
of spatial databases are discussed in \cite{RodBerCani08}.
One of the key criteria to decide about the update to apply is
minimality of geometric changes. Another important criteria may
be the semantics of spatial objects, which makes changes over
the geometry of one type of object more appropriate than
others. For this work, the repair semantics is a rule applied automatically. It assumes that no previous
knowledge about the quality and relevance of geometries exists
and, therefore, it assumes that geometries are all equally important.

On the basis on the minimality condition on geometric changes and
the monotonicity property of some topological predicates  \cite{RodBerCani08}, we propose
to solve inconsistencies with respect to SICs of the form
(\ref{eq:format0}) through shrinking of
geometries. Notice  that this repair semantics will be used as
an instrumental concept to formalize consistent query answers (no actual modification over the database occurs).
As such, it defines what part of the geometry is not in conflict with respect to a set of integrity constraints and can, therefore, be part of a consistent answer.

Shrinking geometries eliminates conflicting parts of geometries without adding new uncertain geometries by enlargement. In this way, we are
considering a proper subset of the possible changes to fix
spatial databases proposed in~\cite{SERUB00}. We disregard
translating objects, because they will carry potentially new
conflicts; and also creating new objects (object splitting),
because we would have to deal with null or unknown thematic
attributes.

The SICs of the form  (\ref{eq:format0}) exclude the topological predicate {\it Disjoint}. The reason
is that falsifying an atom $\nit{DJ}(g_1,g_2)$ by shrinking geometries is not possible, unless we make one
of them empty. However, doing so would heavily depend upon our definition of this topological predicate
for empty regions. Since we opted for not making our approach and results depend on this particular
definition, we prefer to exclude the {\it Disjoint} predicate from our considerations. The study of other  repair semantics that sensibly includes the topological predicate ${\it Disjoint}$ will be left for future work.

Technically, a database $\D$ violates a constraint
$\forall \bar{x}_1 \bar{x}_2 \forall^{\neq \go}s_1 s_2$ $\neg (R_1(\bar{x}_1;$ $
s_1) \wedge R_2(\bar{x}_2 ; s_2) \wedge \varphi \wedge
T(s_1,s_2))$, with $T \in {\cal T}$,\footnote{For simplicity and without lost of
generality, in the examples we consider denial constraints with
at most two spatio-relational predicates and one topological predicate. However, a denial
constraint of the form (\ref{eq:format0}) may have more
spatio-relational predicates and topological predicates.} when there are data values
$\bar{a}_1,\bar{a}_2,g_1,g_2$, with $g_1, g_2 \neq \go$, for the variables in the
constraint such that $(R_1(\bar{x}_1; s_1)$ $\wedge R_2(\bar{x}_2
; s_2) \wedge \varphi \wedge T(s_1,s_2))$ becomes true in the
database under those values. This is denoted with $\D \models
(R_1(\bar{x}_1; s_1) \wedge R_2(\bar{x}_2 ; s_2) \wedge \varphi
\wedge T(s_1,s_2))$ $[\bar{a}_1,\bar{a}_2,g_1,g_2]$. When this
is the case, it is possible to restore consistency of $\D$  by
 shrinking $g_1$ or $g_2$ such that $T(g_1,g_2)$ becomes false.

We can compare geometries, usually an original geometry and its shrunk version, by means of a
distance function that refers to their areas.  We assume that
$\nit{area} \in {\cal O}$ is an operator that computes the area of
a geometry.

\begin{definition} \em
\label{def:functionDistance}  For  regions $g_1,g_2$,
~$\delta(g_1,g_2) ~=~ \nit{area}(\nit{Difference}(g_1,g_2) \cup \nit{Difference}(g_2,g_1))$.
\boxtheorem \end{definition}

\noindent
Since we will compare a region $g_1$ with a region $g_2$
obtained by shrinking $g_1$,  it will hold $\delta(g_1,g_2)
\geq 0$.  Indeed, when comparing $g_2 \subseteq g_1$\footnote{$\subseteq$ stands for geometric inclusion},  the distance function can be simplified by $\delta(g_1,g_2) = \nit{area}(\nit{Difference}(g_1,g_2))$. We  will assume that it is possible to
compare geometries through the distance function by correlating their
tuples, one by one. This requires a
correspondence between instances.

\begin{definition}\em \label{def:link}  Let $\SD, \SD'$ be database
instances of schema $\Sigma$.  $\SD'$  is $(\SD,f_{\SD'})$-{\em indexed} if  $f_{\D'}$ is a bijective function from $\SD$ to $\SD'$, such that, for all $c_1,\ldots, c_n, s$: $f_{\D'}(R(c_1,\ldots,c_n;s)) = R(c_1,\ldots,c_n;s')$, for some region $s'$. \boxtheorem
\end{definition}

\noindent In a $(\SD,f_{\D'})$-indexed instance $\D'$ we can
compare tuples one by one with their counterparts in instance
$\SD$. In particular, we can see how the geometric attribute
values differ. In some cases there is an obvious function
$f_{\D'}$, for example, when there is a key from a subset of
$\mathcal{A}$ to the spatial attribute $S$, or when relations
have a surrogate key for identification of tuples.  In these
cases we simply use the notion of $\SD$-indexed. When the
context is clear, we also use $f$ instead of $f_{\D'}$.

\begin{example} \label{ex:PTA} \em (example \ref{ex:SDB} cont.)
Consider the relational schema ${\it LandP(idl,}$ ${\it name,}$
${\it owner};\textit{geometry)}$. For the instance $\SD$ given
in Example~\ref{ex:SDB}, the following instance $\D'$ is
 $(\nit{\D,f})$-indexed

\begin{center}
{\small
\begin{tabular} {|cccc|} \hline
\multicolumn{4}{|c|}{\textbf{LandP}}\\\hline
{\it idl } & {\it name} &{\it owner} & {\it geometry}\\ \hline
$idl_1$&$n_1$& $o_1$&$g_7$\\
$idl_2$&$n_2$& $o_2$&$g_8$\\
$idl_3$&$n_3$& $o_3$&$g_9$\\
\cline{1-4}
\end{tabular}}
\end{center}

\noindent Here,  $f({\it LandP}({\it idl}_1, n_1, a_1;g_1))$ $
=$ $ {\it LandP}({\it idl}_1, n_1,$ $ a_1; g_7)$, etc.
\boxtheorem \end{example}

\noindent  When restoring consistency,  it may be
necessary to consider different combinations of tuples and SICs.
Eventually, we should obtain a new instance, hopefully
consistent, that we have to compare to the original instance in
terms of their distance.

\begin{definition}\em \label{def:dist}  Let $\SD,\SD'$ be spatial
database instances over the same schema $\Sigma$, with $\SD'$
$(\SD,f)$-indexed. The {\it distance} $\Delta(\SD,\SD')$ ~between
$\SD$ and $\SD'$ ~is the numerical ~value $\Delta(\SD,$
$\SD')= \Sigma_{\bar{t} \in \SD}\delta(\Pi_S(\bar{t}),$
$\Pi_S(f(\bar{t})))$, where $\Pi_S(\bar{t})$ is the projection of
tuple $\bar{t}$ on its spatial attribute $S$. \boxtheorem
\end{definition}

\noindent Now it is possible to define a ``repair semantics", which is
independent of the geometric operators used to shrink
geometries.

\begin{definition} \em \label{def:generalrepair} Let
$\D$ be a spatial database instance over schema $\Sigma$,
$\Psi$ a set of SICs, such that $\D \not \models \Psi$. (a) An
{\em s-repair} of $\SD$ with respect to $\Psi$ is a database
instance $\SD'$ over $\Sigma$, such that: (i) $\SD' \models
\Psi$. (ii) $\SD'$ is $(\SD,f)$-indexed. (iii) For every tuple
$R(c_1,\ldots,c_n;g) \in \SD$, if $f(R(c_1,$ $\ldots,c_n;g))$ $
= R(c_1,\ldots,c_n;g')$, then $g' \subseteq g$. (b) A {\it
minimal s-repair} $\D'$ of $\D$ is a repair of $\D$ such that,
for every repair $\D''$ of $\D$, it holds $\Delta(\D,\D'') \geq
\Delta(\D,\D') $. \boxtheorem \end{definition}

\begin{proposition} \em
If $\SD$ is consistent with respect to $\Psi$, then  $\SD$ is also its only minimal s-repair.
\end{proposition}

\hproof{For  $\SD' = \SD$, it holds: (i) $\SD' \models \Psi$,
(ii) $\SD'$ is $(\SD,f)$-indexed, (iii) for every tuple
$R(c_1,\ldots,c_n;g) \in \SD$, if $f(R(c_1,$ $\ldots,c_n;g))$ $
= R(c_1,\ldots,c_n;g')$, then $g' = g$. In this case,
$\Delta(\D,\D') = 0$.  Any other consistent instance $\SD''$
obtained by shrinking any of $\D$'s geometries and still
obtaining admissible geometries gives $\Delta(\D,\D'') > 0$.}
\noindent This is an ``ideal and natural" repair semantics that
defines a collection of {\em semantic repairs}. The definition
is purely set-theoretic and topological in essence. It is worth
exploring the properties of this semantics and its impact on
properties of consistent query answers (as invariant under
minimal s-repairs) and on logical reasoning about them.
However, for a given database instance we may have a continuum and infinite number
of s-repairs since between two points we have an infinite number of points, which we  want to
avoid for representational and computational reasons.

In this work we will consider an alternative repair semantics
that is more operational in nature (cf. Definition
\ref{def:repair}), leaving the previous one for reference. This
{\em operational} definition of repair  makes it possible to
deal with repairs in current spatial DBMSs and in terms of
standard geometric operators (cf. Lemma \ref{le:lemma1}). Under this definition, there will always
be a  finite number of repairs for a given instance.
Consistency will be restored by applying a finite sequence of
admissible transformation operations to conflicting geometries.

It is easy to see that each true relationship (atom) of the
form $T(g_1,g_2)$, with $T \in {\cal T}$, can be falsified by
applying an admissible transformation in ${\cal O}^a$ to $g_1$ or
$g_2$. Actually, they can be falsified in a {\em canonical
way}. These canonical falsification operations for the
different topological atoms are presented in Table
\ref{ta:admissibleT}. They  have the advantages of: (a) being
defined in terms of the admissible operators,  (b) capturing
the repair process  in terms of the elimination of conflicting
parts of geometries, and (c) changing one of the  geometries participating in a conflict.

\begin{table}[!h]
\begin{center}
\smallskip
\begin{threeparttable}
\begin{tabular}{|p{0.8in} |p{4 in}|} \hline \phantom{x}\hfill
 Pred. $T$~~~~~~~ & \phantom{x} \hfill \centerline{A true atom $T(g_1,g_2)$ becomes a false atom $T(g'_1,g'_2)$ with}\\\hline \hline
 \phantom{x}
 ${\it OV}$ &\phantom{x} 1. ~If $\nit{area}(g_1 \cap g_2) \leq \nit{area}(g_1 \smallsetminus g_2)$:\\
 &$~~~~g'_1 = {\it Difference}(g_1,g_2)$, $g'_2= g_2$.\\
        &2. ~If $\nit{area}(g_1 \cap g_2) > \nit{area}(g_1 \smallsetminus g_2)$:\\
 & $~~~~g'_1 = {\it Difference}(g_1, {\it Difference}(g_1, g_2))$, $g'_2= g_2$. \\
&3.~If $\nit{area}(g_1 \cap g_2) \leq \nit{area}(g_2 \smallsetminus g_1)$:\\
 & $~~~~g'_2 =  {\it Difference}(g_2,g_1)$, $g'_1= g_1$.\\
 &4.~If $\nit{area}(g_1 \cap g_2) > \nit{area}(g_2 \smallsetminus g_1)$:\\
 &  $~~~~g'_2 ={\it Difference}(g_2,{\it Difference}(g_2, g_1))$, $g'_1= g_1$.\\  \hline
  \hline
      \phantom{x}
 ${\it IS,CB}$ &\phantom{x} 1.~If $\nit{area}(g_1 \cap g_2) \leq \nit{area}(g_2 \smallsetminus g_1)$:\\
 & $~~~~g'_2 =  {\it Difference}(g_2,g_1)$, $g'_1= g_1$.\\
 &2.~If $\nit{area}(g_1 \cap g_2) > \nit{area}(g_2 \smallsetminus g_1)$:\\
 &  $~~~~g'_2 ={\it Difference}(g_2,{\it Difference}(g_2, g_1))$, $g'_1= g_1$.\\
 & 3.~$g'_1 = {\it Difference}(g_1,g_2)$, $g'_2= g_2$. \\\hline
  \hline
        \phantom{x}
 ${\it IC,CV}$ &\phantom{x} 1. ~If $\nit{area}(g_1 \cap g_2) \leq \nit{area}(g_1 \smallsetminus g_2)$:\\
 &$~~~~g'_1 = {\it Difference}(g_1,g_2)$, $g'_2= g_2$.\\
 & 2. ~If $\nit{area}(g_1 \cap g_2) > \nit{area}(g_1 \smallsetminus g_2)$:\\
 & $~~~~g'_1 = {\it Difference}(g_1, {\it Difference}(g_1, g_2))$, $g'_2= g_2$. \\
&3.~$g'_2 =  {\it Difference}(g_2,g_1)$, $g'_1= g_1$.\\  \hline \hline
 \phantom{x}  ${\nit II,WI,CO}$&
\phantom{x} 1.~$g'_1 = {\it Difference}(g_1, g_2)$, $g'_2= g_2$.   \\
&2.~$g'_2 = {\it Difference}(g_2, g_1)$, $g'_1= g_1$.\\
\hline      \hline
\phantom{x} ${\nit TO, IT }$ & \phantom{x}1.~$g'_1 = {\it Difference}(g_1, {\it buffer}(g_2,d))$, $g'_2= g_2$. \\
&2.~$g'_2 = {\it Difference}(g_2, {\it buffer}(g_1,d))$, $g'_1= g_1$.\\ &\small{(See Remark~\ref{re:d} for definition of $d$)}\\
\hline \hline
\phantom{x} ${\it EQ}$  & \phantom{x} 1.~$g'_1 = \go$, $g'_2= g_2$.\\ &2.~$g'_2 = \go$, $g'_1= g_1$. \\ \hline
\end{tabular}
\end{threeparttable}
\caption{Admissible transformations}\label{ta:admissibleT}
\end{center}
\end{table}

More specifically, in Table \ref{ta:admissibleT} we indicate,
for each relation $T \in {\cal T}$, alternative  operations
that falsify a true atom of the form $T(g_1,g_2)$. Each of them
makes changes on one of the geometries, leaving the other
geometry unchanged. The list of  {\em  canonical
transformations} in this table prescribes particular
 ways of applying the admissible operators of Definition
\ref{de:spatialop}. Later on, they will also become the {\em
admissible or legal} ways of transforming geometries with the
purpose of restoring consistency.

For example, Table \ref{ta:admissibleT} shows that for
$\nit{Overlaps} (OV)$, there are in principle four ways to make
false an atom ${\it Overlaps}(g_1,g_2)$ that is true. These are
the alternatives 1. to 4. in that entry, where alternatives 1.
and 2. change geometry $g_1$; and alternatives 3. and 4. change geometry $g_2$.
Only one of these alternatives that satisfies its
condition is expected to be chosen to falsify the atom.  A
minimal way to change a geometry depends on the relative size
between overlapping and non-overlapping areas: (i) when the
overlapping area between $g_1$ and $g_2$ is smaller than or equal to
their non-overlapping areas, a minimal change over
geometry $g_1$ is ${\it Difference}(g_1, g_2)$, and over $g_2$
is  ${\it Difference}(g_2, g_1)$ (cases 1. and 3. for $OV$ in
Table~\ref{ta:admissibleT}). (ii) When the non-overlapping
areas of $g_1$ or $g_2$ are smaller than the overlapping area,
a minimal change over geometry $g_1$ is ${\it Difference}(g_1,$
${\it Difference}(g_1, g_2))$, and over geometry $g_2$ is ${\it
Difference}(g_2,$ ${\it Difference}(g_2, g_1))$ (cases 2. and
4. for $OV$ in Table~\ref{ta:admissibleT}).

For the case when ${\nit Equals}(g_1,g_2)$ is true, the transformations in Table~\ref{ta:admissibleT} make either geometry, $g_1$ or $g_2$ empty to falsify the atom. However, there are other alternatives that  by shrinking geometries would achieve the same result, but also producing smaller changes in terms of the affected area. A natural candidate update consists in applying the transformation $g'_1 =\textit{Difference}(g_1,\textit{Buffer}(\textit{Boundary}(g_2),d))$ (similarly and alternatively for $g_2$). In this case, we just  take away from $g_1$ the part of the internal area of width $d$ surrounding the boundary of $g_1$, to make it different from $g_2$. We did not follow this alternative for practical reasons: having two geometries that are topologically equal could, in many cases, be the result of duplicate data, and one
of them should be eliminated.  Moreover, this alternative, in comparison with the officially adopted in this work, may create new conflicts with respect to other SICs. Avoiding them whenever possible will be  used later, when designing a polynomial algorithm for CQA based on the core of an inconsistent database instance (see Section~\ref{sec:core}).

Table~\ref{ta:admissibleT}  shows that {\it Touches} and
{\it Intersects} are predicates for which the eliminated area
is not completely delimited by the real boundary of objects.
Actually, we need to separate the touching boundaries. We
do so by buffering a distance $d$ around one of the geometries
and taking the overlapping part from the other one.\footnote{The buffer operator does not introduce new points in the geometric representation of objects, but it translates the boundary a distance $d$ outwards.}

The following result is obtained directly from Table \ref{ta:admissibleT}.

\begin{lemma} \label{le:lemma1}\em For each topological predicate $T \in {\cal T}$ and true ground atom $T(g_1,g_2)$, there are
geometries $g_1',g_2'$ obtained by means of the corresponding
admissible transformation in Table   \ref{ta:admissibleT}, such
that $T(g_1',g_2')$ becomes false.\boxtheorem
\end{lemma}
The following definition defines, for each geometric predicate $T$, a binary geometric operator $tr^T$ such that, if $T(g_1,g_2)$ is true, then $tr^T(g_1,g_2)$ returns a geometry $g_1'$
such that $T(g_1',g_2)$ becomes false. The definition is based on the transformations that affect geometry $g_1$ in
Table \ref{ta:admissibleT}.

\begin{definition} \label{def:ad}\em Let  $T \in {\cal T}$ be a topological predicate. We define an admissible transformation operator $\nit{tr}^T:  \nit{Ad}
\times \nit{Ad} \to \nit{Ad}$ as follows:
\begin{itemize}
\item[(a)] If $T(g_1,g_2)$ is false, then
    $\nit{tr}^{T}(g_1,g_2) := g_1$.
\item [(b)] If $T(g_1,g_2)$ is true,  then:
\begin{eqnarray*}
tr^{T}(g_1,g_2) & :=  & \left \{ \begin{array}{l}
\nit{Difference}(g_1,g_2) \textrm{ if } \nit{area}(g_1 \cap g_2) \leq \nit{area}(g_1 \smallsetminus g_2)\\
\nit{Difference}(g_1, \nit{Difference}(g_1,g_2)), \textrm{ otherwise }
\end{array} \right. \\
&& \textrm{ for } T \in  \{OV,IC,CV\}; \\
tr^{T}(g_1,g_2) & := & \nit{Difference}(g_1,g_2), \textrm{ for } T \in \{IS,CB,II,WI,CO\}; \\
tr^{T}(g_1,g_2) & := & \nit{Difference}(g_1,\nit{Buffer}(g_2,d)) \textrm{ for } T \in \{TO,IT\}; \\
tr^{T}(g_1,g_2) & := & \go \textrm{ for } T \in \{EQ\}. \hspace{8.5cm} \Box
\end{eqnarray*}
\end{itemize}
\end{definition}
It can be easily verified that the admissible operations
$\nit{tr}^T$, applied to admissible geometries, produce admissible
geometries. They can be seen as macros defined in terms of the
basic operations in Definition \ref{de:spatialop}, and inspired
by Table \ref{ta:admissibleT}.  The idea is that the operator
$tr^{T}$ takes $(g_1,g_2)$, for which $T(g_1,g_2)$ is true, and
makes the latter false by transforming $g_1$ into $g'_1$, i.e.,
$T(g'_1,g_2)$ becomes false.

Definition \ref{def:ad} can also be used to formalize the transformations
on geometry $g_2$ indicated in Table \ref{ta:admissibleT}. First,
notice that for  the converse
predicate $T^c$ of predicate $T$ it holds: $T^c(g_1,g_2)$ true iff $T(g_2,g_1)$.
Secondly,
 the converse of a transformation  operator
can be defined by ~$(\nit{tr}^T)^c := \nit{tr}^{(T^c)}$.
In consequence, we can apply
$\nit{tr}^{T^c}$ to $(g_2,g_1)$, obtaining the desired transformation of geometry $g_2$.
In this way,  all the cases in Table
\ref{ta:admissibleT} are covered. For example, if we
want to make false a true atom ${\it Inside}(g_1,g_2)$, we can
apply $tr^{IS}\!(g_1,g_2)$, but also $tr^{IC}\!(g_2,g_1)$.

\begin{example}\em  Table ~\ref{tab:figTransform}  illustrates the application
of the admissible transformations to restore consistency of
predicates $T \in \{{\nit Overlaps},
{\nit Touches}\}$. The dashed boundary is the result of
applying {\it Buffer}$(g,d)$. \boxtheorem
\end{example}

\begin{table}[!h]
\begin{center}
\begin{tabular}{lp{3cm}p{3cm}p{3cm}p{3cm}}
\hline
\\
{\bf $T(g_1,g_2)$}&{\bf Original}&$Tr^T(g_1,g_2)$&$Tr^T(g_2,g_1)$\\
\hline
\\
${\nit OV}$&\includegraphics[width=1cm]
{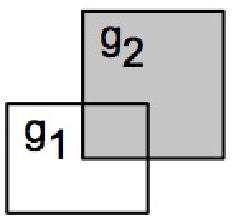}&\includegraphics[width=1 cm]
{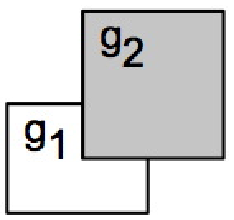}&\includegraphics[width=1 cm]{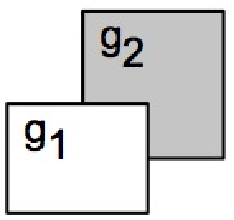}\\
\cline{2-4}
&
\vspace{0.05 mm}
\includegraphics[width=1cm] {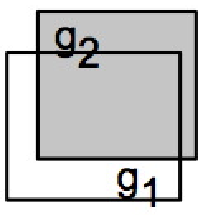}&\vspace{0.05 mm}\includegraphics[width=1 cm] {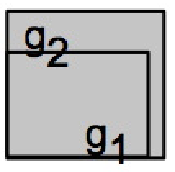}&\vspace{0.05 mm}\includegraphics[width=1 cm]{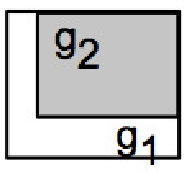}\\
\hline
\\


${\nit TO}$& \includegraphics[width=1.5 cm] {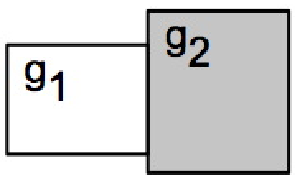}&\includegraphics[width=1.8 cm] {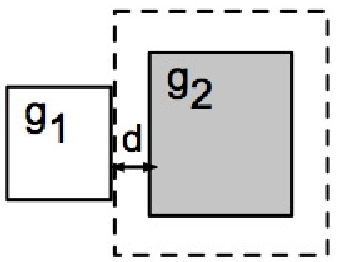}&\includegraphics[width=1.8 cm] {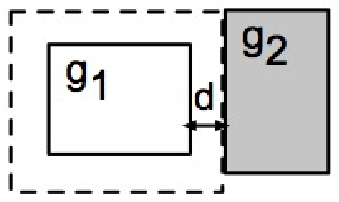}\\
\hline
\end{tabular}
\caption{Examples of admissible
transformations}\label{tab:figTransform}
\end{center}
\end{table}

\noindent We now define the notion of \textit{accessible instance} that
results from an original instance, after applying admissible
transformation operations to geometries. The application of
sequences of operators solves violations of SICs. Accordingly,
the accessible instances are defined by induction.

\begin{definition}\label{def:instances}\em
Let $\D$ be a database instance. $\D'$ is an {\it accessible
instance} from $\D$ (with respect to a finite set of SICs $\Psi$), if $\D'$  is obtained after
applying, a finite number of times, the following inductive
rules (any of them, when applicable):\\
(1). ~ $\D' = \D$.

\noindent
(2).~ There is an accessible instance $\D_0$ from $\D$,
    such that, for some $\psi \in \Psi$  with a topological
    predicate $T$, $\D_0 \not \models \psi$\footnote{$\psi$ may have more than one topological predicate.} through tuples
    $R_1(\bar{a}_1;g_1)$ and $R_2(\bar{a}_2;g_2)$ in $\D_0$, for which
    $T(g_1,g_2)$ is true; and\\
\hspace*{1cm}(a) $\D' = \D_0 \smallsetminus
\{R_1(\bar{a_1},g_1)\}
\cup \{R_1(\bar{a_1},\nit{tr}^{\!T}\!(g_1,g_2)) \}$,~ or\\
\hspace*{1cm}(b) $\D' = \D_0 \smallsetminus
\{R_2(\bar{a_2},g_2)\} \cup \{
R_2(\bar{a_2},\nit{tr}^{\!T^c}\!\!(g_2,g_1)) \}$.\boxtheorem
\end{definition}

\begin{figure}[t]
\begin{center}
\begin{multicols}{2}
{\small \begin{tabular} {|cccc|} \hline
\multicolumn{4}{|c|}{\textbf{LandP}}\\\hline
{\it idl} & {\it name} & {\it owner} & {\it geometry}\\ \hline
$idl_1$&$n_1$&$o_1$&$g_1$\\
$idl_2$&$n_2$&$o_2$&$g_2$\\
$idl_3$&$n_3$&$o_3$&$g_3$\\
\cline{1-4}
\end{tabular}}
\includegraphics[width=2.8 cm] {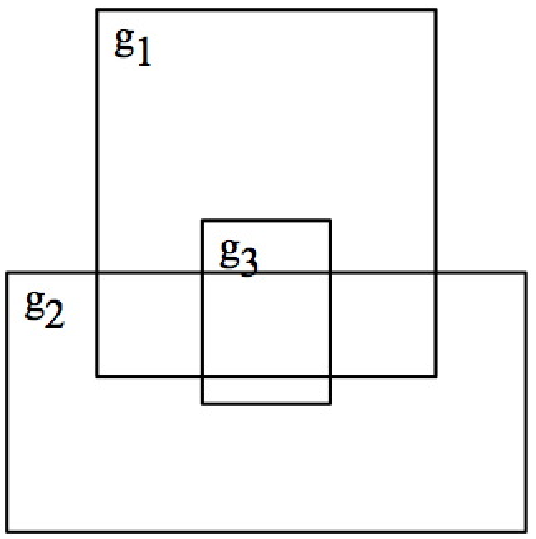}
\end{multicols}

(a)

\begin{multicols}{2}
{\small \begin{tabular} {|cccc|} \hline
\multicolumn{4}{|c|}{\textbf{LandP}}\\\hline {\it idl} & {\it
name} & {\it owner} & {\it geometry}\\ \hline
$idl_1$&$n_1$&$o_1$&$g_1'$\\
$idl_2$&$n_2$&$o_2$&$g_2$\\
$idl_3$&$n_3$&$o_3$&$g_3$\\
\cline{1-4}
\end{tabular}}
\includegraphics[width=2.8cm] {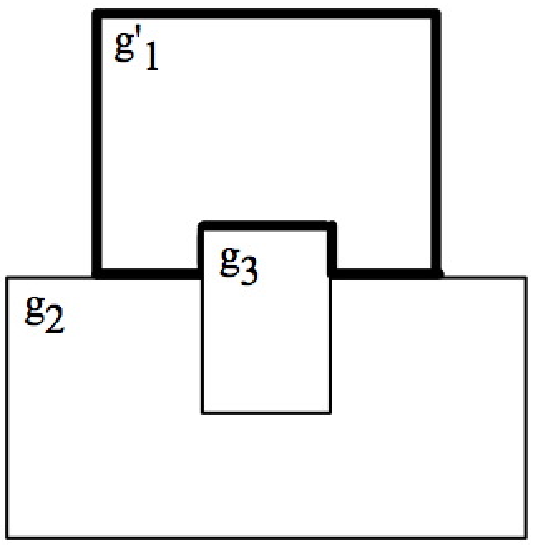}
\end{multicols}

(b) \caption{An accessible instance: (a) original
instance, and (b) accessible transformation (geometry with thick
boundary changed)} \label{fig:induction}
\end{center}
\end{figure}

\begin{example}\em
Consider the database instance in Figure~\ref{fig:induction}(a)
that is inconsistent with respect to SIC~(\ref{eq:constraint}).
An accessible instance from this inconsistent database is in
Figure~\ref{fig:induction}(b), where only $g_1$ has changed.
This can be expressed in the following way:
$LandP(idl_1,n_1,o_1;g_1') =
LandP(idl_1,n_1,o_1;\nit{tr}^{II}(\nit{tr}^{II}(g_1,g_2),g_3))$.
\boxtheorem\end{example}

\noindent Given a database $\D$, possibly inconsistent, we are
interested in those accessible instances $\D'$ that are
consistent, i.e., $\D' \models \Psi$. Even more, having the
repairs in mind, we have to make sure that admissible instances
from $\D$ can still be indexed with $\D$.

\begin{proposition}\label{def:index}\em Let $\D'$ be an accessible instance
from $\D$. Then, $\D'$ is $f$-indexed to $\D$ via an
index function $f$,  that can be defined by induction on
$\D'$.
\end{proposition}

\hproof{To simplify the presentation, we will assume that $\D$
has an index (or surrogate key) $i_0$, that is a one-to-one
mapping from $\D$ to an initial segment $[1,N]$ of
$\mathbb{N}$. Let $\D'$ be an accessible instance from $\D$. We
define $i_{\D'}(R(\bar{a};g)) \in \mathbb{N}$ for tuples in
$\D'$ by induction on $\D'$:\\
(1).~If $\D' = \D$ and $R(\bar{a};g) \in \D$,
    $i_{\D'}(R(\bar{a};g)) = i_0(R(\bar{a};g))$.

\noindent
(2).~If there is an accessible instance $\D_0$ from $\D$ and
$\D_0 \not \models \psi \in \Psi$ through the atoms
$R_1(\bar{a}_1;g_1)$, $R_2(\bar{a}_2;g_2)$, and $T(g_1,g_2)$
with $T$ and $T^c$ the converse relation of
$T$:\\
\begin{tabular}{p{0.05 cm}p{15.95 cm}}
&(a) ~$\D' = \D_0 \smallsetminus \{R_1(\bar{a}_1,g_1)\} \cup \{
R_1(\bar{a}_1,tr^T(g_1,g_2)) \}$, and
$i_{\D'}(R_1(\bar{a}_1;tr^T(g_1,g_2)))$ $ =
i_{\D_0}(R_1(\bar{a}_1,g_1))$  and $i_{\D'}(R_2(\bar{a}_2;g_2))
= i_{\D_0}(R_2(\bar{a}_2,g_2))$, or\\
&(b)~ $\D' = \D_0 \smallsetminus \{R_2(\bar{a}_2,g_2)\} \cup \{
R_2(\bar{a}_2,tr^{T^c}(g_2,g_1)) \}$, and
$i_{\D'}(R_1(\bar{a}_1;g_1)) = i_{\D_0}(R_1(\bar{a}_1,g_1))$
and $i_{\D'}(R_2(\bar{a}_2;tr^{T^c}(g_2,g_1))) =
i_{\D_0}(R_2(\bar{a}_2,g_2))$.}
\end{tabular}
\noindent Any two accessible instances $\D'$ and  $\D''$ can be
indexed via $\D$ in a natural way, and thus, they can be
compared tuple by tuple. In the following, we will assume, when
comparing any two accessible instances in this way, that there
is such an underlying index function $f$. Now we give the
definition of operational repair.

\begin{definition} \em \label{def:repair}  Let
$\D$ be an instance over schema $\Sigma$ and $\Psi$ a finite set of
SICs. (a) An {\em o-repair} of $\SD$ with respect to $\Psi$ is
an instance $\D'$ that is accessible from $\SD$, such that $\D'
\models \Psi$.  (b) A {\it minimal o-repair} $\D'$ of $\D$ is
an o-repair of $\D$ such that, for every o-repair $\D''$ of
$\D$, $\Delta(\D,\D'') \geq \Delta(\D,\D') $. (c)
$\nit{Rep}(\SD,\Psi)$ denotes the set of minimal o-repairs of
$\SD$ with respect to $\Psi$. \boxtheorem
\end{definition}

\noindent The distances $\Delta(\D,\D'')$ and $\Delta(\D,\D')$
in this definition are relative to the corresponding
index functions, whose existence is guaranteed by
Proposition~\ref{def:index}.  Unless otherwise stated,  this is
the repair semantics we refer to in the remainder of the paper, in
particular, in the definition of consistent query answer in Section \ref{se:CQA}. In consequence,
in the following  a repair is an o-repair, and the same applies
to minimal repairs. Even more, whenever we refer to repairs, we
should understand that minimal repairs are intended.

\begin{figure}[t!]
\begin{center}
\begin{multicols}{3}
{\small
\begin{tabular} {|p{0.2cm}p{0.5cm}p{0.6cm}c|} \hline
\multicolumn{4}{|c|}{\textbf{LandP}}\\\hline
{\it idl} & {\it name} & {\it owner} & {\it geometry}\\ \hline
$idl_1$&$n_1$&$o_1$&$g_1$\\
$idl_2$&$n_2$&$o_2$&$g_2$\\
$idl_3$&$n_3$&$o_3$&$g_3$\\
$idl_4$&$n_4$&$o_4$&$g_4$\\
\cline{1-4}
\end{tabular}

~~~~~

\begin{tabular} {|cc|} \hline
\multicolumn{2}{|c|}{\textbf{Building}}\\\hline
{\it idb} & {\it geometry}\\ \hline
$idb_1$&$g_5$\\
$idb_2$&$g_6$\\
\cline{1-2}
\end{tabular}}

\includegraphics[width=3cm] {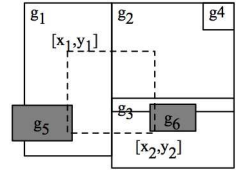}
\end{multicols}
\caption{Inconsistent database instance } \label{inconEx}
\end{center}
\end{figure}

\begin{figure}[t!]
\begin{center}
\begin{tabular}{cp{0.3 cm}c}
\includegraphics[width=3cm] {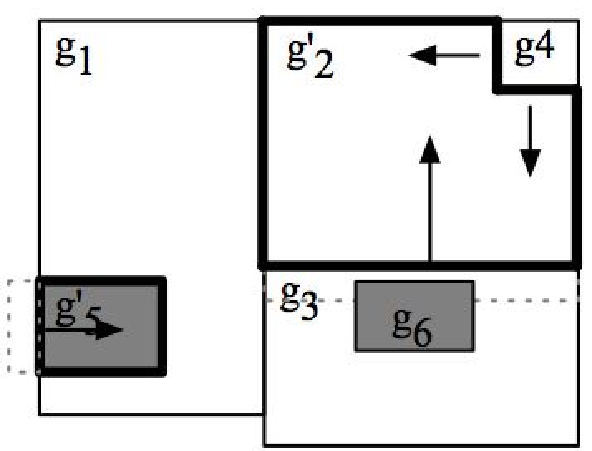}&&\includegraphics[width=3cm]
{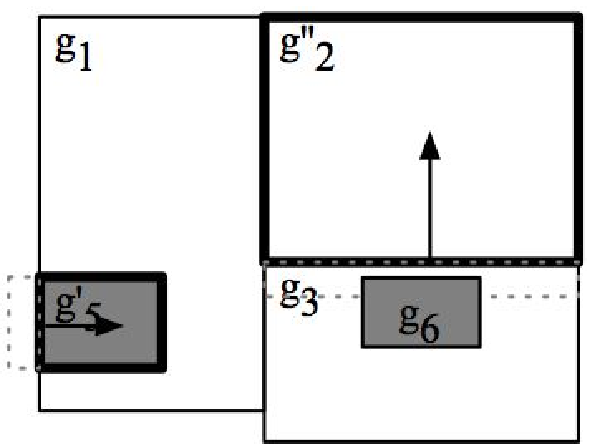}\\
(a)&&(b)\\
\end{tabular}
\caption{Minimal repairs} \label{fig:repairs}
\end{center} \end{figure}

\begin{example} \em \label{ex:PTA2} Consider database schema in Example \ref{ex:SDB}. The
instance $\D$ in Figure~\ref{inconEx} is inconsistent with
respect to the SICs (\ref{eq:constraint}) and (\ref{eq:build}),
because the land parcels with geometries $g_2$ and $g_3$
overlap, and so do the land parcels with geometries $g_2$ and
$g_4$. Likewise, buildings with geometry $g_5$ and $g_6$
\textit{partially} overlap land parcels with geometries $g_1$ and $g_2$,
respectively.

Figure~\ref{fig:repairs} shows the two minimal repairs of
$\SD$. In them, the regions with thicker boundaries are the
regions that have their geometries changed. For  the minimal
repair in Figure \ref{fig:repairs}(a), the inconsistency
involving geometries $g_2$ and $g_3$ is repaired by applying
${\it Difference(g_2,g_3)}$ to $g_2$, i.e., removing from $g_2$
the whole overlapping geometry, and keeping the geometry of
$g_3$ as originally. Notice that due to the interaction between
integrity constraints, if we apply ${\it Difference(g_3,g_2)}$
to $g_3$, i.e., we remove the whole overlapping area from $g_3$,
we still have an inconsistency, because the building with
geometry $g_6$ will continue partially overlapping geometries
$g_2$ and $g_3$.  Thus, this change will require an additional
transformation to ensure that $g_6$ is completely covered or
inside of $g_3$.

In the same minimal repair (Figure \ref{fig:repairs}(a)),  the
inconsistency between $g_2$ and $g_4$ is repaired by shrinking
$g_2$, eliminating its area that overlaps $g_4$. This is obtained by applying ${\it
Difference(g_2,g_4)}$ to $g_2$.  Finally, the inconsistency
between $g_1$ and $g_5$ is repaired by removing from $g_5$ its part
that does not overlap with geometry
$g_1$. In principle, we could have repaired this inconsistency by eliminating the
overlapping region between $g_1$ and $g_5$, but this is not a
minimal change.

In the second minimal repair (Figure \ref{fig:repairs}(b)),
geometries $g_2$ and $g_5$  undergo the same changes than
those in the first minimal repair (Figure \ref{fig:repairs}(a)),
but  the inconsistency between $g_2$ and $g_4$ is restored by
eliminating geometry $g_4$, i.e., applying ${\it Difference}$
$(g_4,g_2)= \go$. \boxtheorem \end{example}

\noindent Notice that, by applying admissible transformation
operators to restore consistency, the whole part of a geometry
that is in conflict with respect to another geometry is
removed. In consequence, given that there are finitely many
geometries in the database instance and finitely many SICs, a
finite number of applications of admissible transformations are
sufficient to restore consistency.  This contrasts with the
s-repair semantics, which can yield even a continuum of
possible consistency-restoration transformations. Keeping the
number of repairs finite may be crucial for certain mechanisms
for computing  consistent query answers, as those  as we will
show in the next sections. Actually, we will use existing
geometric operators as implemented in spatial DBMSs in order to
capture and compute the consistency-restoring geometric
transformations. This will be eventually used to obtain
consistent query answers for an interesting class of spatial
queries and SICs in Section~\ref{sec:computingCore}.

Despite the advantages of using o-repairs, the following
example shows that an o-repair may not be minimal under the
s-repair semantics.

\begin{example} \em \label{ex:SORepairs} The
instance $\D$ in Figure~\ref{sosemantics} is inconsistent with
respect to the SICs~(\ref{eq:constraint}) and (\ref{eq:build}),
because the land parcels with geometries $g_1$ and $g_2$
internally intersect and buildings with geometry $g_3$ and
$g_4$  overlap land parcels
with geometries $g_1$ and $g_2$, respectively.

Figures~\ref{fig:sorepairs}(a) and (b) show the  minimal s-repair (Definition~\ref{def:generalrepair}) and o-repairs of
$\SD$ (Definition~\ref{def:repair}), respectively. In them, the regions with thicker boundaries are the
regions that had their geometries changed. Here, by applying s-repair semantics we obtain one minimal
repair (Figure~\ref{fig:sorepairs}(a)) that takes the partial
conflicting parts from both land parcels $g_1$ and $g_2$ in
conflict, and leave unchanged the geometries of buildings $g_3$
and $g_4$. Instead, for the o-repair semantics, each repair
takes the whole conflicting parts from one of the land parcels
$g_1$ or $g_2$ in order to satisfy SIC (\ref{eq:constraint}),
and to satisfy SIC (\ref{eq:build}), each repair eliminates the conflict between the new version of
$g_1$ and building $g_3$ or  between the new version of $g_2$ and
building $g_4$. This makes up
to four possible o-repairs (Figure~\ref{fig:sorepairs}(b)),
which are not minimal with respect to the single s-repair.
\boxtheorem
\end{example}

\begin{figure}[t]
\begin{center}
\begin{multicols}{3}
{\small \begin{tabular} {|p{0.2cm}p{0.5cm}p{0.6cm}c|} \hline
\multicolumn{4}{|c|}{\textbf{LandP}}\\\hline
{\it idl} & {\it name} & {\it owner} & {\it geometry}\\ \hline
$idl_1$&$n_1$&$o_1$&$g_1$\\
$idl_2$&$n_2$&$o_2$&$g_2$\\
\cline{1-4}
\end{tabular}

~~~~~~
\begin{tabular} {|cc|} \hline
\multicolumn{2}{|c|}{\textbf{Building}}\\\hline
{\it idb} & {\it geometry}\\ \hline
$idb_1$&$g_3$\\
$idb_2$&$g_4$\\
\cline{1-2}
\end{tabular}}

\includegraphics[width=3cm] {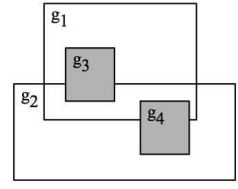}
\end{multicols}
\caption{An inconsistent database instance} \label{sosemantics}
\end{center}
\end{figure}

\begin{figure}[t!]
\begin{center}
\begin{tabular}{cp{0.3 cm}c}
\includegraphics[width=2cm] {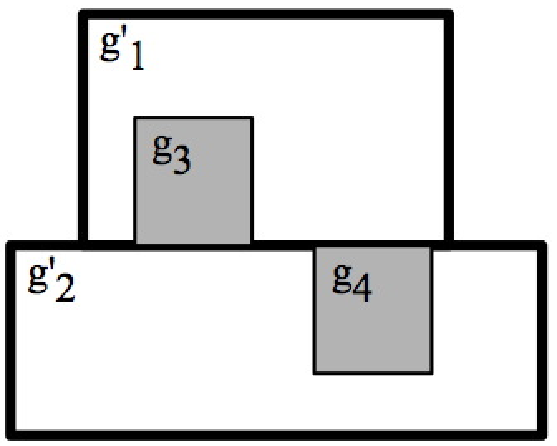}&&\includegraphics[width=8cm]
{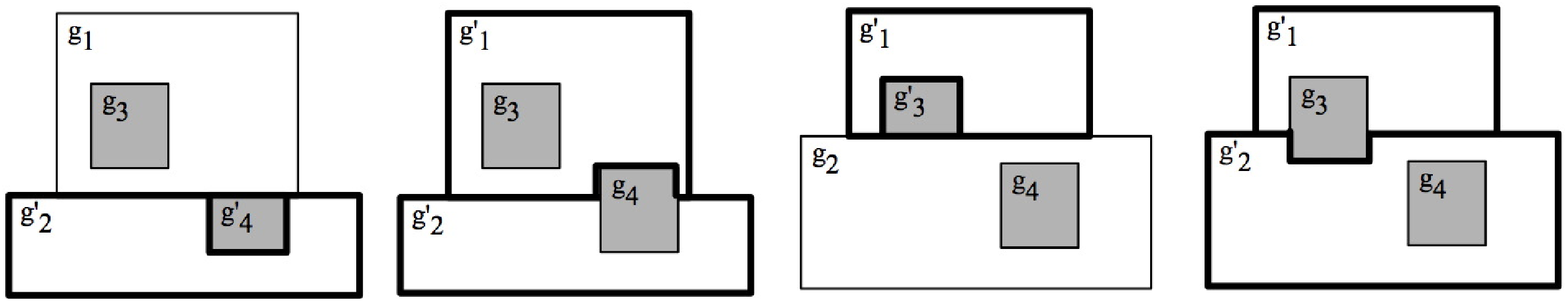}\\
(a)&&(b)\\
\end{tabular}
\end{center}
\caption{Minimal repairs: (a) minimal s-repair  and (b) minimal o-repairs (thick boundaries show geometries that have changed)}\label{fig:sorepairs}
 \end{figure}

\noindent S-repairs may take away only parts of a
geometry that participate in a conflict. On the other side, they do not force a conflicting
geometry to become empty in cases where o-repairs would do so. For instance, consider  a true atom
$\nit{Equals}(g_1,g_2)$ that has to be falsified. A s-repair
can be obtained by shrinking one of the two geometries just a
little, without making it empty. However, by using
admissible transformations, we can only falsify this atom by
making one of the geometries empty. In this case, a minimal o-repair is not a minimal
s-repair.

\begin{proposition} \label{prop:properties} \em  Let $\D$ be a
database instance, and $\Psi$ a set of SICs. Then the following properties for o-repairs hold: (a) If $\D$ is
consistent with respect to $\Psi$, then $\D$ is its only minimal
o-repair. (b) If $\D'$ is an
$(\SD,f)$-indexed o-repair of $\D$ and
$f(R(\bar{a};g)) = R(\bar{a};g')$, then $g' \subseteq g$.
(c) The set of o-repair for $\D$ is finite and non-empty.
\end{proposition}

\hproof{(a) By the inductive definition of o-repair, an
    admissible transformation operator is applied to a
    geometry $g$ when $g$ is in conflict with other
    geometry $g'$ in $\D$. Since a consistent database
    instance does not contain conflicting tuples, none of
    the transformations operators is applicable and the
    consistent database instance is its only o-repair.\\
(b) The application of each admissible
    transformation $\nit{tr}^T(g_1,g_2)$, with $T \in
    \cal{T}$, has five possible outcomes:  $g_1$,  $\go$,
    $\nit{Difference}(g_1,g_2)$, $\nit{Difference}(g_1,\nit{Difference}(g_1,$ $g_2))$, or
    $\nit{Difference}(g_1,\nit{Buffer}(g_2,d))$.
    Then, by definition of operator
    $\nit{Difference}$ (cf. Definition~\ref{de:spatialop}),
    $\nit{tr}^T(g_1,g_2) \subseteq g_1$.\\
    (c)   $\D$ has a finite number $N$ of tuples; and $\Psi$, a
    finite number of integrity constraints. In consequence, there is a finite number of conflicts, i.e., sets of
    tuples that simultaneously participate in the violation of one element $\psi$ of  $\Psi$ via their geometries.
    Each of these conflicts are solved by shrinking some of those geometries.  Each application of an operator $O$, chosen for a finite set of them, according to the inductive definition of o-repair solves an existing conflict by falsifying at least one of the $\mathcal{T}$-atoms in a ground instance of $\psi$.
    In principle, the application of such an operator $O$ may produce new conflicts; however, it strictly decreases the total  geometrical area  of the database instance. More precisely, if
    $A(D') := \Sigma_{R(\bar{t};g) \in D'} \nit{area}(g)$, then $A(D') > A(O(D'))$, where $O(D')$ is the instance resulting from the conflict resolving operator $O$ to $D'$. In particular, $A_0$ is the area $A(D)$ of the original instance $D$.

    Now we reason by induction on the structure of o-repairs. The application of a one-conflict solving operator $O$ to an instance $D_{n-1}$ produces an instance $D_n$ with
     $A(D_n) < A(D_{n-1})$. Moreover,
    $A(D_{n-1}) - A(D_n) > \epsilon > 0$, where $\epsilon$  represents a lower bound of the area reduction at each inductive step.

We claim that, due to our repair semantics, this lower bound $\epsilon$ depends on the initial instance $D$, and not on $n$. In order to prove this, let us first remark that an admissible  region is fully determined by its boundaries.  Now we  prove that the regions in any accessible instance depend on the regions in the original database instance, or, more precisely, by the boundaries delimiting those regions. We prove it by induction on the number of inductive steps of the definition of accessible instances.

First, we prove that it works for the first repair transformation on the original database instance. Let $g_1'= Tr^T(g_1,g_2)$  be the first transformation applied on  region $g_1$ to create the accessible instance $\D_1$ from the  original database instance $\D$. For  $T \in {\cal T} \setminus \{{\it TO,IT}\}$, and following the definitions of admissible transformations in Table~\ref{ta:admissibleT},  the geometry $g_1'$ is either $\go$ or a region whose boundary  is formed by parts of the boundaries that limit regions $g_1$ and  $g_2$  (see example of overlapping regions in Figure~\ref{fig:exampleBoundaries}). For  $T \in  \{{\it TO,IT}\}$,  $g'_1$  is formed by  parts of the boundary of region $g_1$ and the boundary created by buffering $d$ around $g_2$.   So, in this case,
the boundary of $g_1'$ depends exclusively on the boundaries of $g_1$ and $g_2$, and of the constant $d$.

\begin{figure}[t!]
\begin{center}
\begin{tabular}{c}
\includegraphics[width=6cm] {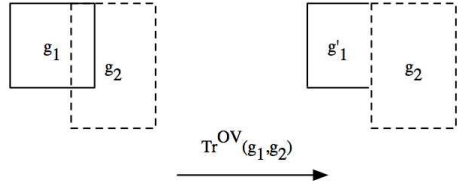}
\end{tabular}
\end{center}
\caption{Example of a region's boundary after geometric transformation}\label{fig:exampleBoundaries}
 \end{figure}

Let assume that the geometries in an accessible instance $\D_n$ obtained after $n$ inductive steps are $\go$ or regions whose boundaries depend on the original instance.  In the $n+1$ inductive step, another transformation $g_i'= Tr^{T'}(g_i,g_j)$ is applied. Following the definition of admissible transformations, $g_i'$ becomes the geometry $\go$ or a region whose boundary is formed by part of the boundary of $g_i$ and part of the boundary of $g_j$, as in the first inductive step. Thus, $g_i'$ also depends on the original database instance. This establishes our claim.

Now, by the Archimedean property of real numbers, there is a number $M$ such that $A_0 - M\epsilon < 0$. Thus, after
 a finite number of iterations (i.e. applications of conflict-solving operators), we reach a consistent instance
  or an instance with area $0$, i.e. all of whose geometries are empty, which would be consistent too.

  Notice that the
  number $M$ provides an upper bound on the number of times we can apply operators to produce
a repair. At each point we have a finite number of choices. So, the overall number of o-repairs that can be produced is finite.}


\noindent The following example shows
that, even when applying admissible transformations, there may
be exponentially many minimal repairs in the size of the
database, a phenomenon already observed with relational repairs
with respect to functional dependencies \cite{BC03}.

\vspace{1mm}
\begin{example} \em \label{ex:PTA3} Consider the schema in
Example \ref{ex:SDB}, and the  SIC (\ref{eq:constraint}). The
database instance contains $n$ spatial tuples, as shown in
Figure~\ref{fig:noverlap}. There are $n\!-\!1$ overlappings and $n$ overlapping
geometries.

\begin{figure}[h!]
\begin{center}
\includegraphics[width= 5 cm] {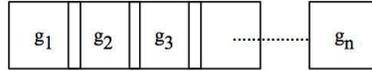}
\caption{Exponential number of repairs} \label{fig:noverlap}
\end{center} \end{figure}

\noindent In order to solve each of those overlaps, we have the
options of shrinking either one of the two regions involved. We
have $2^{n-1}$ possible minimal repairs. \boxtheorem
\end{example}
The following remark is important when estimating the data
complexity of repairs, because, in this case, data complexity does not only depend on the
number of tuples, but also on the size of geometric representations.

\begin{remark} \em Transformation operators that make
geometries empty reduce the size of geometric
representations. Any other admissible transformation operator
$\nit{tr}^T(g_1,$ $g_2)$ shrinks $g_1$, and uses $g_1$ and $g_2$  to define the new
boundary of $g_1$. Thus, we are using, in a simple manner, the original geometric representation
(e.g. points in the boundaries of the original
geometries) to define a new geometry.  It is clearly the case that there is a polynomial
upper bound on the size of the representation of a new geometry in an o-repair in terms of the
size of the original database, including representations of geometric regions.
\boxtheorem
\end{remark}

\section{Consistent  Query Answers} \label{se:CQA}
We can use the concept of minimal repairs as an auxiliary concept to define, and possibly compute, consistent answers to a relevant class of
queries in ${\cal L}(\Sigma)$.

A general conjunctive query  is of the form:
\begin{equation}\label{eq:query}
{\cal Q}(\bar{v}):~ \exists \bar{y} (R_1(\bar{x}_1;s_1)\wedge
\cdots \wedge R_n(\bar{x}_n;s_n) \wedge \varphi),
\end{equation}
where $\bar{v} = (\bigcup_i(\bar{x}_i \cup \{s_i\})) \smallsetminus \bar{y}$ are the free variables, and  $\varphi$ is a conjunction of built-in atoms over thematic attributes or  over spatial attributes that involve
topological predicates in $\cal T$ and geometric operators in $\cal O$. We also add a {\em safety condition}, requiring that variables in $\varphi$ also appear in some of the $R_i$.
For example, the following is a conjunctive  query:
\begin{eqnarray*}
{\cal Q}(x,y;s): &&\exists s_1 s_2(R(x;s_1) \wedge
R(y;s_2)  \wedge \nit{Intersects}(s_1,s_2) \wedge  x \neq y  \wedge s =
\nit{Difference}(s_1,s_2)).
\end{eqnarray*}
We will consider the simpler but common and  relevant class of conjunctive queries that are  {\em operator free}, i.e., conjunctive queries of the form (\ref{eq:query}) where $\varphi$ does not contain geometric operators. We will also study in more detail two particular classes of conjunctive queries:
\begin{itemize}
\item [(a)] {\it Spatial range queries} are of the form
\begin{equation} \label{eq:range}
{\cal Q}(\bar{u};s): \exists \bar{z} (R(\bar{x};s) \wedge T(s,w)),
\end{equation} with  $w$ a spatial constant, and $\bar{z} \subseteq \bar{x}$.
This is a  ``query window'', and its free variables are
those in $\bar{u} = (\bar{x} \smallsetminus \bar{z})$ or in
$\{s\}$.

\item [(b)] {\it Spatial join queries} are of the form
\begin{equation} \label{eq:join}
{\cal Q}(\bar{u};s_1,s_2): \exists \bar{z} (R_1(\bar{x}_1;s_1) \wedge R_2(\bar{x}_2;s_2)\wedge T(s_1,s_2)),
\end{equation}
with $T \in \cal{T}$, and $\bar{z} \subseteq \bar{x}_1 \cup
\bar{x}_2$. The free variables are those in $\bar{u}
=((\bar{x}_1 \cup \bar{x}_2) \smallsetminus \bar{z})$ or in
$\{s_1, s_2\}$.
\end{itemize} We call {\em basic conjunctive queries} to queries of the form~(\ref{eq:range}) or (\ref{eq:join}) with ${\it T} \in \{{\it IIntersects, Intersects}\}$.

\begin{remark} \label{rem:queries} \em
Notice that for these two classes of queries we project on all
the geometric attributes. We will also assume that the free
variables correspond to a set of attributes of $R$ with its
key of the form (\ref{eq:primarykey}). More precisely, for
range queries, the attributes associated with $\bar{u}$ contain
the key of $R$. For join queries, $\bar{u} \cap \bar{x_1}$ and
$\bar{u} \cap \bar{x_2}$ contain the key for relations $R_1$,
$R_2$, respectively. This is a common situation in spatial
databases, where a geometry is retrieved together with its key
values. \boxtheorem
\end{remark}

\noindent Given a query ${\cal Q}(\bar{x};\bar{s})$, with free
thematic variables $\bar{x}$ and free geometric variables
$\bar{s}$, a sequence of thematic/spatial constants
$\langle\bar{c}; \bar{g}\rangle$ is an answer to the query  in
instance $\D$ if and only if $\SD \models {\cal Q}(\bar{c};
\bar{g})$, that is the query $\cal Q$ becomes true in $\D$ as a
formula when its free variables $\bar{x}, \bar{s}$ are replaced
by the constants in $\bar{c}, \bar{g}$, respectively.  We
denote with ${\cal Q}(\D)$ the set of answers to ${\cal Q}$ in
instance $\D$.

\begin{example}
\em \label{ex:CQA}   \ Figure \ref{queryEx} shows an instance for the
schema ${\cal R} = \{{\it LandP(idl};$ ${\it geometry)}$, ${\it
Building(idb;}$ ${\it geometry)}\}$. Here, $\nit{idl, idb}$ are keys for their relations. Dark rectangles represent
buildings, and white rectangles represent land parcels. The
queries ${\cal Q}_1$ and ${\cal Q}_2$ below  are a  range and a
join query, respectively. For the former, the spatial constant
is the spatial window shown in Figure~\ref{queryEx}, namely the
(closed) polygon obtained by joining the four points in order
indicated in the query.
\begin{eqnarray}
{\cal Q}_1({\nit idb;g})&:&~ \nit{Building}(idb;g)~
\wedge \nonumber \\
&&{\nit Intersects(g},([x_1,y_1],[x_2,y_1],[x_2,y_2],[x_1,y_2],[x_1,y_1])).\nonumber\\
{\cal Q}_2({\nit idl,idl';g,g'})&:&~ \nit{LandP}(idb;g)~\wedge~\nit{LandP}(idb';g')~\wedge 
{\nit Touches}(g,g').\nonumber
\end{eqnarray}

\begin{figure}[t!]
\begin{multicols}{3}
{\small
\begin{tabular} {|cc|} \hline
\multicolumn{2}{|c|}{\textbf{LandP}}\\\hline
{\it idl} & {\it geometry}\\ \hline
${\it idl}_1$&$g_1$\\
${\it idl}_2$&$g_2$\\
${\it idl}_3$&$g_3$\\
\cline{1-2}
\end{tabular}

\begin{tabular} {|cc|} \hline
\multicolumn{2}{|c|}{\textbf{Building}}\\\hline
{\it idb} & {\it geometry}\\ \hline
${\it idb}_1$&$g_4$\\
${\it idb}_2$&$g_5$\\
\cline{1-2}
\end{tabular}}

\includegraphics[width=3cm] {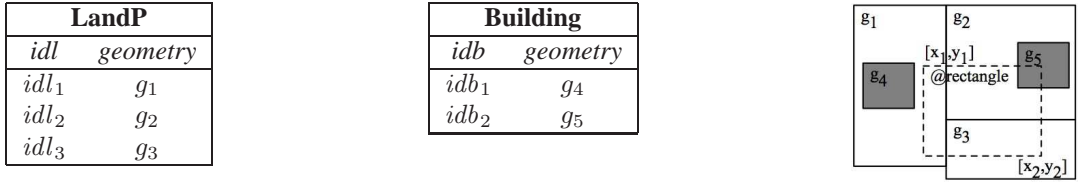}
\end{multicols}
\caption{~Example of a range query}
\label{queryEx}
\end{figure}

\noindent The answer to ${\cal Q}_1$  is $\langle{\nit
idb}_2;g_5\rangle$. The answers to ${\cal Q}_2$ are:
$\{\langle{\nit idl}_1,{\nit idl}_2;g_1,g_2\rangle,$
$\langle{\nit idl}_2,{\nit idl}_3;g_2,g_3\rangle,$~
$\langle{\nit idl}_1,{\nit idl}_3;$ $g_1,g_3\rangle$,~
$\langle{\nit idl}_2,{\nit idl}_1;g_2,g_1\rangle,$
$\langle{\nit idl}_3,{\nit idl}_2;g_3,g_2\rangle,$
$\langle{\nit idl}_3,{\nit idl}_1;g_3,g_1\rangle\}$.
\boxtheorem \end{example} 

\noindent Now we define the notion of consistent answer to a
conjunctive query.

\begin{definition}
\em \label{de:consistent-answers} Consider an instance $\SD$, a
set $\Psi$ of SICs, and a conjunctive query ${\cal
Q}(\bar{x};\bar{s})$. A tuple of thematic/geometric constants
$\langle c_1, \ldots$ $, c_m; g_1, \ldots, g_l\rangle$ is a
{\em consistent answer} to ${\cal Q}$ with respect to $\Psi$
if:~ (a) For every $\SD' \in \nit{Rep}(\SD,\Psi)$, there exist
$g_1', \ldots, g_l'$ such that $\SD' \models {\cal Q}(c_1,
\dots, c_m; g_1',\ldots,g_l')$. (b) $g_i$ is the intersection
over all regions $g_i'$ that satisfy (a) and are correlated to
the same tuple in $\D$.\footnote{Via the correlation function
$f$, cf. Definition \ref{def:link}.}~
 ${\nit Con}({\cal Q},\D,\Psi)$ denotes the set of consistent answers
to  ${\cal Q}$ in instance $\D$ with respect to $\Psi$.
\boxtheorem \end{definition}

\noindent Since ${\cal Q}$ is operator free, the
regions $g_i'$ appear in  relations of the repairs, and then
$f^{-1}$ can be applied. However, due to the intersection of geometries, the
geometries in a consistent answer may not belong to the
original instance or to any of its repairs.

In contrast to the definition of consistent answer to
a relational query \cite{ABC99}, where a consistent answer is an answer in every repair, here we have an aggregation of query answers via the
geometric intersection and grouped-by thematic tuples. This definition is similar in spirit to consistent answers to aggregate relational queries with group-by \cite{ABC03,caniupan07,fuxSigmod05}.

This definition of consistent
 answer allows us to obtain more significative answers than in the relational case,  because
when shrinking geometries, we cannot expect to have, for a
fixed tuple of thematic attribute values, the same geometry in
every repair. If we did not use the intersection of geometries, we
might lose or not have consistent answers due to the lack of
geometries in common among repairs.

\begin{figure}[t]
\begin{center}
\begin{tabular}{c p{0.4 cm} c}
\begin{tabular}{|cc|} \hline
\textit{idl}&\textit{geometry}\\
\hline
$\nit{idl}_1$&$g_1$\\
$\nit{idl}_2$&$g'_2$\\
$\nit{idl}_3$&$g_3$\\
\cline{1-2}

\end {tabular}&&
\begin{tabular}{c}
\includegraphics[width=2.5cm]{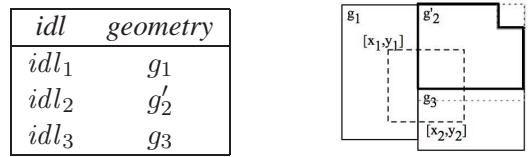}
\end{tabular}
\end{tabular}
\caption{~Consistent answers} \label{fig:CQALand}
\end{center}
\end{figure}

\begin{example} \em  (example \ref{ex:PTA2} cont.) \label{ex:CQA2}
Consider the spatial range query\\ ${\cal Q}(\nit{idl;geometry}):$~
$\exists \nit{name~ owner}
(\nit{LandP(idl,name,owner;geometry)}~ \wedge$\\
\hspace*{3cm}$\nit{Intersects(geometry},([x_1,y_1],[x_2,y_1],[x_2,y_2],[x_1,y_2],[x_1,y_1])),$\\
which is expressed in the SQL language as:
\begin{eqnarray}
&\textrm{SELECT }& {\it idl, geometry} \nonumber \\
&\textrm{FROM  }&{\it LandP} \nonumber \\
&\textrm{WHERE  }&{\it Intersects}({\it
geometry},([x_1,y_1],[x_2,y_1],[x_2,y_2], [x_1,y_2],[x_1,y_1])).\nonumber
\end{eqnarray}
Now, consider the two minimal repairs in Figure~\ref{fig:repairs}.
In them, objects ${\it idl_1}$ and ${\nit idl}_3$  do not
change geometries, whereas object ${\it idl_2}$ does, from
$g_2$ to $g'_2, g_2''$, resp. (cf. Figure \ref{fig:repairs}(a), (b), resp.).

From the first repair we get the following (usual) answers to
the query: $\langle \nit{idl}_1; g_1\rangle$, $\langle
\nit{idl}_2; g_2'\rangle, \langle \nit{idl}_3; g_3\rangle$.
From the second repair, we obtain   $\langle \nit{idl}_1;
g_1\rangle, \langle \nit{idl}_2;g_2''\rangle, \langle
\nit{idl}_3;$ $g_3\rangle$. The consistent answers are the
tuples shown in Figure~\ref{fig:CQALand}, where the answers
obtained in the repairs are grouped by an {\it idl} in common,
and the associated geometries are intersected. In this figure,
the geometry with thicker lines corresponds to the intersection
of geometries obtained from different repairs.

From a practical point of view, the consistent query answer
could include additional information about the degree in which
geometries  differ from their corresponding original
geometries. For example, for the answer  $\langle \nit{idl}_2;
g_2'\rangle$, an additional information could be the relative
difference between areas $g_2$ and $g_2'$, which is calculated
by  $\delta(g_2,g_2')/area(g_2)$.\boxtheorem \end{example}

\section {Core-Based  CQA}\label{sec:core}
The definition of consistent query answer relies on the
auxiliary notion of minimal repair. However, since we may have
a large number of repairs, computing consistent answers by
computing, materializing, and finally querying all the repairs
must be avoided whenever there are more efficient mechanisms at hand. Along these lines, in this section we present a
methodology for  computing consistent query answers to a
subclass of conjunctive queries with respect to certain kind of SICs. It works in polynomial time (in data complexity), and
does not require the explicit computation of the database
repairs.

We start by defining the  ${\it core}$, which is a single
database instance associated with the class of repairs. We will use the
core to consistently answer a subclass of conjunctive queries.
Intuitively, the core is the ``geometric intersection" of the
repairs, which is obtained by intersecting the geometries in
the different repair instances that correlate to the same
thematic tuple.

\begin{definition} \em \label{def:core} For an instance $\D$
and a set $\Psi$ of SICs, the {\em core} of $\D$ is the
instance $\D^\star$ given by~ $\D^*:=$ $\{R(\bar{a};g^\star)~|~
R \in {\cal R}, \mbox{ there is } R(\bar{a};g) \in \D \mbox{
and } g^\star = \bigcap \{g'~|$ $R(\bar{a}; g') \in \D' ~\mbox{
for some } ~\D' \in \nit{Rep}(\SD,\Psi)$ ~$\mbox{and }
~R(\bar{a};g') =f_{\D'}(R(\bar{a};g))\}\}$. Here, $f_{\D'}$ is
the correlation function for $\D'$.\footnote{Here, $\bigcap$ is a set-theoretic intersection of geometries.} \boxtheorem
\end{definition}

\noindent Sometimes we will refer to $\D^\star$  by $\bigcap^g
\nit{Rep}(\SD,\Psi)$. However, it cannot be understood as the
set-theoretic intersection of the repairs of $\D$. Rather it is
a form of geometric intersection of geometries belonging to different repairs and grouped by common thematic attributes. It is important to remark
that the keys of relations remain in the repairs, and therefore
they appear in the core of a dimension instance.

\begin{example} \em \label{ex:core} Figure \ref{fig:core} shows the {\em core}
of  the database instance in Figure \ref{inconEx} considering
the repairs in Figure~\ref{fig:repairs}. Here, $g^*_2$ results
from the geometric intersection of geometries $g_2'$ and
$g_2''$ of the minimal repairs in Figure~\ref{fig:repairs}.
Similarly, $g^*_5$ is  $g_5'$, because the latter is shared by
both minimal repairs in Figure~\ref{fig:repairs}. Geometry
$g_4$ becomes $\go$ in the core.  All other geometries in the
core are unchanged with respect to geometries in the original
database instance. \boxtheorem \end{example}

\begin{figure}[t]
\begin{center}
\begin{multicols}{3}
\begin{tabular} {|cccc|} \hline
\multicolumn{4}{|c|}{\textbf{LandP$^*$}}\\\hline
$idl_1$&$n_1$&$o_1$&$g^*_1$\\
$idl_2$&$n_2$&$o_2$&$g^*_2$\\
$idl_3$&$n_3$&$o_3$&$g^*_3$\\
$idl_4$&$n_4$&$o_4$&$\go$\\
\cline{1-4}
\end{tabular}

\begin{tabular} {|cc|} \hline
\multicolumn{2}{|c|}{\textbf{Building$^*$}}\\\hline
$idb_1$&$g^*_5$\\
$idb_2$&$g^*_6$\\
\cline{1-2}
\end{tabular}

\includegraphics[width=3cm] {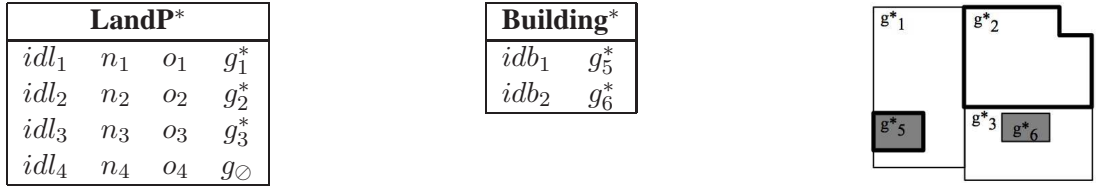}
\end{multicols} \end{center}
\caption{~The core of an instance}
\label{fig:core}
\end{figure}

\noindent Notice the resemblance between the definitions of
consistent answer and the core. Actually, it is easy to see
that $\D^\star = \bigcup_{\ R \in \cal{R}}{\nit Con}({\cal
Q}_R,\D,\Psi)$, where the query ${\cal Q}_R(\bar{x};s):
R(\bar{x};s)$ asks for the tuples in relation $R$.

The  ${\it core}$ is defined as the geometric intersection of
all database repairs. However, as we will show, for a subset of SICs  we can actually determine the ${\it core}$ without
computing these repairs. This is possible for SICs of the form:
\begin{equation}
\forall \bar{x}_1\bar{x}_2s_1s_2
\neg(R(\bar{x}_1;s_1) ~\wedge~
R(\bar{x}_2;s_2) ~\wedge~ \bar{x}'_1 \neq \bar{x}'_2
 \wedge~ \it{T}(s_1,s_2)),\label{eq:phicore}
\end{equation}
where ${\it T} \in \{{\it IIntersects, Intersects, Equals}\}$,
$\bar{x}'_1 \subseteq \bar{x}_1$, $\bar{x}'_2 \subseteq
\bar{x}_2$, and both $\bar{x}'_1$ and $\bar{x}'_2$ are the key of $R$. In these SICs there are two occurrences of the same database predicate in the same SIC. The following
example illustrates this class of SICs.

\begin{example} \em  \label{ex:core2}
For the schema ${\cal R} = \{{\it County(idc,}$ ${\it name}$;
${\it geometry)}$, ${\it Lake(idl;}$ ${\it geometry)}\}$, with  ${\nit idc}$ the key of  ${\nit County}$ and ${\nit idl}$ the key of ${\nit Lake}$,
the
following SICs are  of the  form~(\ref{eq:phicore}):
\begin{eqnarray}
\overline{\forall}\neg(\nit{County}(idc_{1},n_{1};s_1) \wedge
\nit{County}(idc_{2},n_{2};s_2) ~\wedge idc_{1} \neq idc_{2}\wedge \nit{IIntersects}(s_1,s_2)). \label{eq:phicore1}
\end{eqnarray}
\begin{equation}
\overline{\forall}\neg(\nit{Lake}(idl_{1};s_1) \wedge
\nit{Lake}(idl_{2};s_2) ~\wedge idl_{1} \neq idl_{2}\wedge \nit{Intersects}(s_1,s_2)). \label{eq:phicore2}
\end{equation}
\boxtheorem
\end{example}

\begin{remark}\label{re:properties}\em
\noindent This subset of SICs has the following
properties, which will be useful when trying to compute the repairs and the core:

\begin{itemize}
\item [(i)] Two SICs of the form~(\ref{eq:phicore}) over
    the same database predicate are redundant due to the
    semantic interrelation of the topological predicates
    {\it IIntersects}, {\it Intersects,} and {\it Equals}: only the constraint that contains the weakest  topological predicate has to be considered.
    For example, {\it Intersects} is weaker  than {\it IIntersects}, and {\it IIntersects} is weaker than {\it Equals}.

\item [(ii)] Conflicts between tuples with respect to SICs
    of the form~(\ref{eq:phicore}) are determined by the
    intersection of their geometries. The conflict
    between two tuples $R(\bar{a}_1;g_1)$ and
    $R(\bar{a}_2;g_2)$ is solved by
    applying a single admissible transformation operator
    $tr^{T}(g_1,g_2)$ (or $tr^{T^c}(g_2,g_1)$) that
    modifies $g_1$ (or $g_2$), and makes $T(g_1,g_2)$ (and
    $T^c(g_2,g_1)$) false.

\item [(iii)]  Solving conflicts with respect to a SIC of
    the form~(\ref{eq:phicore}) is independent from solving a conflict with respect to another SIC of
    form~(\ref{eq:phicore}) over a different database
    predicate.

\item [(iv)] Solving a conflict between two tuples with
    respect to a SIC of the form~(\ref{eq:phicore}) does
    not introduce new conflicts. This is due to the
    definition of admissible transformations and the
    monotonicity property of predicates {\it IIntersects} and {\it Intersects}, which prevent  a shrunk geometry
    (or even an empty geometry) from participating  in a new conflict with an existing geometry in the database (cf. Example~\ref{ex:core3}).

\item [(v)] For any two geometries $g_1$ and $g_2$ in
    conflict with respect to a SIC of the
    form~(\ref{eq:phicore}), there always exist two
    repairs, one with the shrunk version of $g_1$, and
    another  with the shrunk version of $g_2$. This
    guarantees that there exists a minimal repair that contains a
    minimum version of a geometry whose its geometric
    intersections with original geometries in conflict have
    been eliminated (cf. Lemma \ref{prop:all}).
    As a consequence, the core can be computed by
    taking from a geometry all its intersections with other
    geometries in conflict,  disregarding the  order in
    which these intersections are eliminated.

This property is not guaranteed for other kinds of SICs. For instance, consider  Example~\ref{ex:PTA2} with the instance in Figure~\ref{inconEx} and its
corresponding repairs in Figure~\ref{fig:repairs}. Although $g_6$ was originally in conflict with respect to $g_2$, there is no minimal repair where geometry $g_6$
is shrunk.
\boxtheorem
\end{itemize} \end{remark}
We illustrate some of these properties  with the following
example.

\begin{figure}[t]
\begin{center}
\begin{multicols}{3}
{\small
\begin{tabular} {|p{0.4cm}p{0.5cm}c|} \hline
\multicolumn{3}{|c|}{\textbf{County}}\\\hline
{\it idl} & {\it name} &  {\it geometry}\\ \hline
$idc_1$&$n_1$&$g_1$\\
$idc_2$&$n_2$&$g_2$\\
$idc_3$&$n_3$&$g_3$\\
$idc_4$&$n_4$&$g_4$\\
$idc_4$&$n_5$&$g_5$\\
\cline{1-3}
\end{tabular}
}

{\small
\begin{tabular} {|p{0.25cm}c|} \hline
\multicolumn{2}{|c|}{\textbf{Lake}}\\ \hline
{\it idl} & {\it geometry}\\
\hline
$idl_1$&$g_6$\\
$idl_2$&$g_7$\\
 & \\
\cline{1-2}
\end{tabular}
}
\includegraphics[width=4cm]{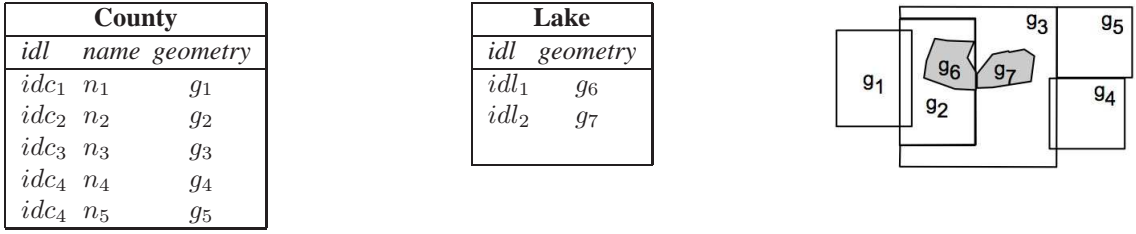}
\end{multicols}
\end{center}
\caption{~An inconsistent database with SICs of the
form~(\ref{eq:phicore})} \label{ex:coreInstance}
\end{figure}

\begin{example} \em \label{ex:core3}(example~\ref{ex:core2}
cont.) Consider the inconsistent instance in Figure
\ref{ex:coreInstance}. In it, counties with
geometries $g_1$, $g_2$ and $g_3$ are inconsistent with
respect to SIC~(\ref{eq:phicore1}), because they internally
intersect. Also, county $g_3$ internally intersects with geometry $g_4$. Lakes with geometries $g_6$ and $g_7$ violate SIC~(\ref{eq:phicore2}), because they intersect (actually they touch).

Conflicts  with respect to SICs (\ref{eq:phicore1}) and (\ref{eq:phicore2}) can be solved in an independent way, since
they do not share predicates (cf. Remark~\ref{re:properties}(iii)). To obtain a repair, consider first SIC~(\ref{eq:phicore1}) and
the conflict between $g_2$ and $g_3$, which is solved by applying
$tr^{{\it II}}(g_2,g_3)$ or $tr^{{\it II}}(g_3,g_2)$.  Any
of these  alternative transformations do not produce
geometries that could  be in conflict with other geometries
unless they were originally in conflict (cf. Remark~\ref{re:properties}(iv)). For instance, if we apply $tr^{{\it II}}(g_3,g_2)$ we obtain a new geometry $g'_3$ that will be in conflict with geometries $g_1$ and $g_4$. These conflicts are not new, since $g_3$ was originally in conflict with these two geometries. Even more, by shrinking $g_2$ or $g_3$, none of the modified geometries could be in conflict with $g_5$. In
addition, although by making $g'_3 = tr^{{\it II}}(g_3,g_2)$
we also solve the conflict between $g_1$ and $g_3$, this is
only accomplished due to the fact the conflicting part of $g_3$ and
$g_1$ has been already eliminated from $g'_3$ (cf. Remark~\ref{re:properties}(v)).

Figure~\ref{coreRepair} shows the sixteen possible minimal repairs that are obtained by considering the eight possible ways in which conflicts with respect to SIC~(\ref{eq:phicore1}) are solved, in combination with the two possible ways in which conflicts with respect to SIC (\ref{eq:phicore2}) are solved.  In this figure thick boundaries represent geometries that have changed. Notice that in this figure we only show $g'_{2_1}$ and not $g'_{2_2}$, since the later corresponds to the empty geometry which is then omitted in the corresponding repairs. The core for this database instance is shown in Figure~\ref{fig:core2}. \boxtheorem
\end{example}

\begin{figure}[t]
\begin{center}
\begin{tabular}{cccc}
\includegraphics[width=3cm]{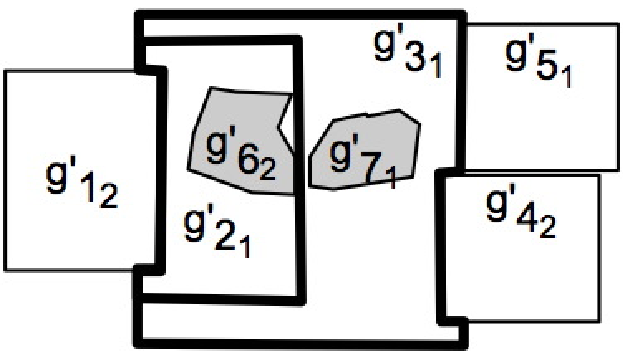}&\includegraphics[width=3cm]{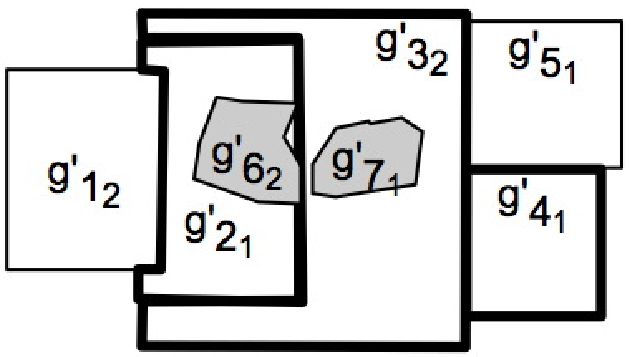}&\includegraphics[width=3cm]{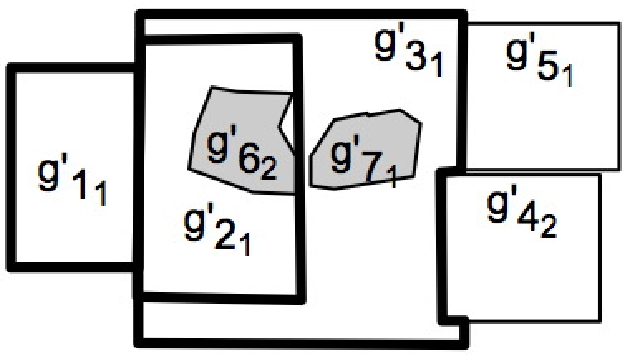}&\includegraphics[width=3cm]{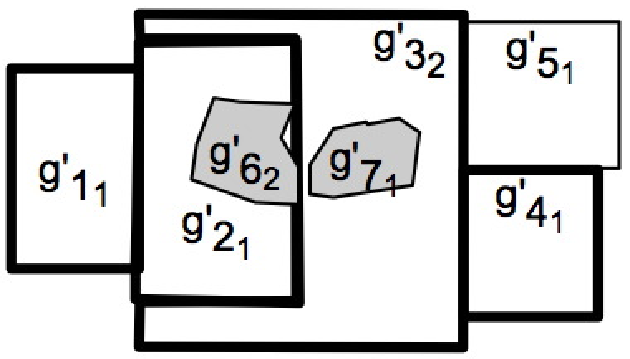}\\
\includegraphics[width=3cm]{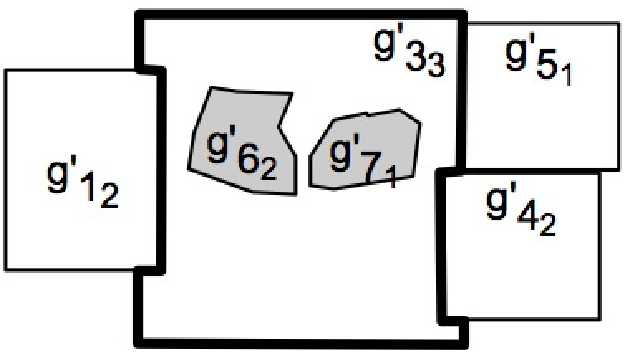}&\includegraphics[width=3cm]{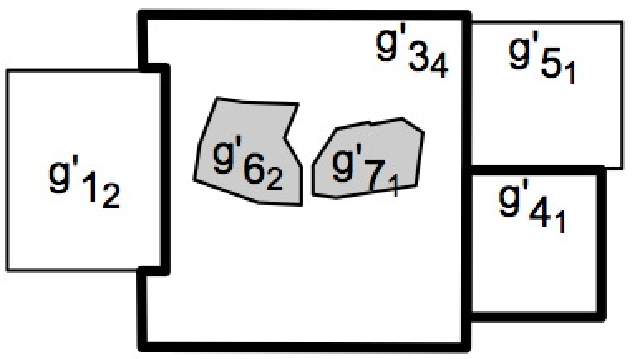}&\includegraphics[width=3cm]{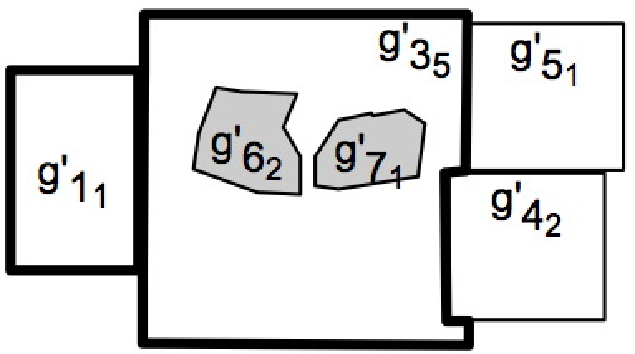}&\includegraphics[width=3cm]{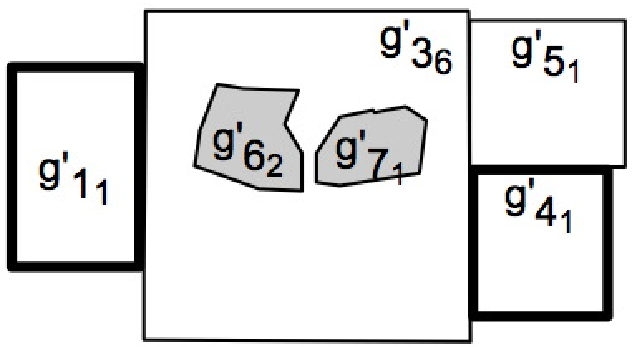}\\
\includegraphics[width=3cm]{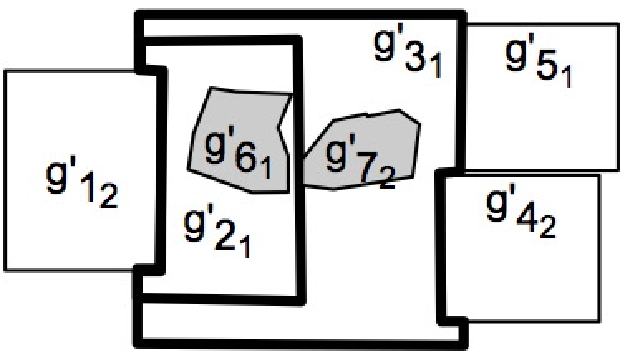}&\includegraphics[width=3cm]{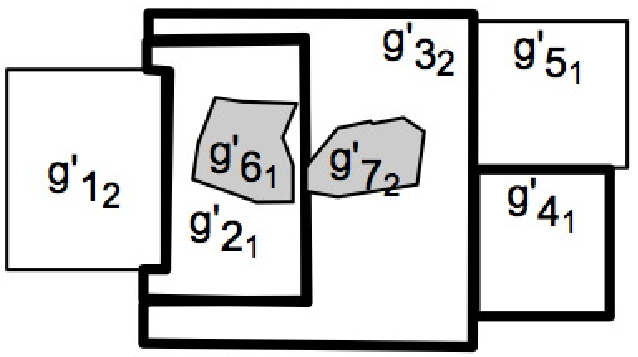}&\includegraphics[width=3cm]{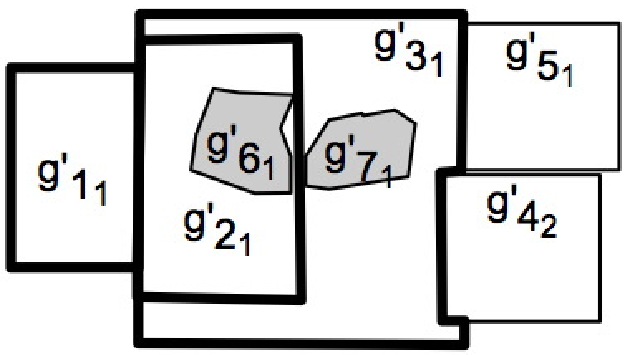}&\includegraphics[width=3cm]{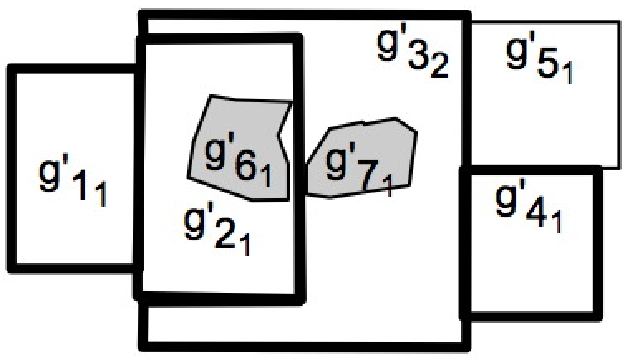}\\
\includegraphics[width=3cm]{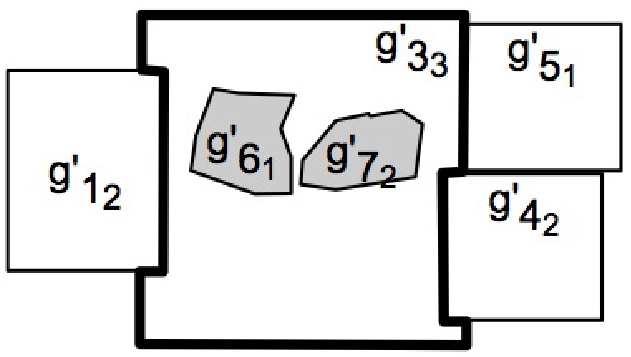}&\includegraphics[width=3cm]{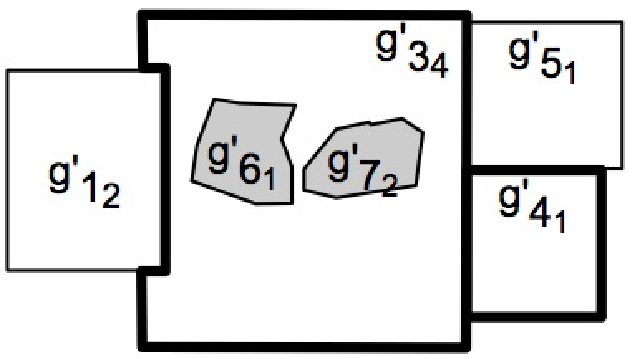}&\includegraphics[width=3cm]{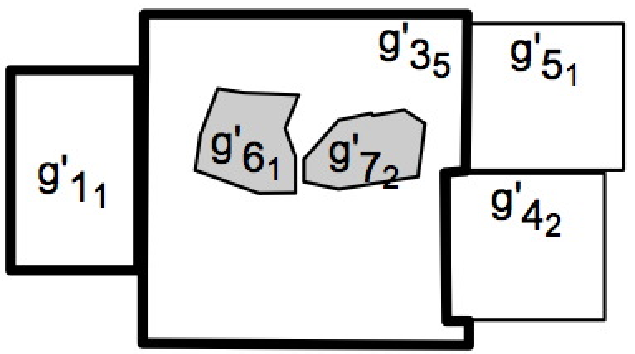}&\includegraphics[width=3cm]{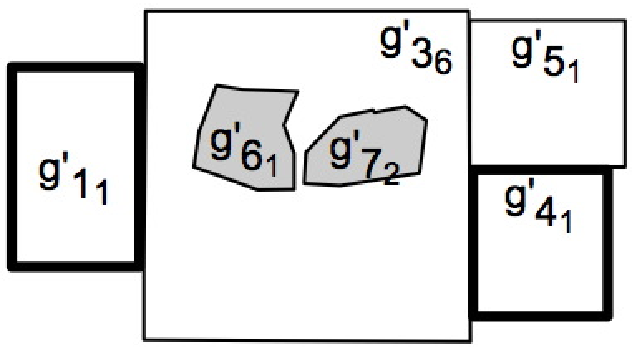}
\end{tabular}
\caption{The sixteen possible repairs of Example~\ref{ex:core3}}
\label{coreRepair}
\end{center}
\end{figure}

\begin{figure}[t]
\begin{center}
\begin{multicols}{3}
{\small
\begin{tabular} {|p{0.4cm}p{0.5cm}c|} \hline
\multicolumn{3}{|c|}{\textbf{County$^*$}}\\\hline
{\it idl} & {\it name} &  {\it geometry}\\ \hline
$idc_1$&$n_1$&$g'_{1_1}$\\
$idc_2$&$n_2$&$\go$\\
$idc_3$&$n_3$&$g'_{3_1}$\\
$idc_4$&$n_4$&$g'_{4_1}$\\
$idc_4$&$n_5$&$g'_{5_1}$\\
\cline{1-3}
\end{tabular}
}

{\small
\begin{tabular} {|p{0.25cm}c|} \hline
\multicolumn{2}{|c|}{\textbf{Lake$^*$}}\\ \hline
{\it idl} & {\it geometry}\\
\hline
$idl_1$&$g'_{6_1}$\\
$idl_2$&$g'_{7_1}$\\
 & \\
\cline{1-2}
\end{tabular}
}
\includegraphics[width=4cm]{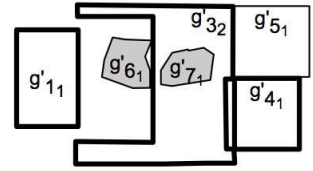}
\end{multicols}
\end{center}
\caption{The core of Example~\ref{ex:core3}}
\label{fig:core2}
\end{figure}

\noindent It is possible to use a tree to represent all the versions that a geometry $g$ may take in the repairs. The root of that tree is the original geometry
$g$, the leaves are all the possible versions of the $g$ in the minimal repairs. The
internal nodes represent partial transformations applied to $g$ as different conflicts in which $g$ participates are solved. For illustration,
Figure~\ref{treeRepair} shows the tree that represents the possible different versions of $g_3$ in the minimal repairs for the inconsistent  instance
in Figure \ref{ex:coreInstance}. Notice that a leaf in this tree represents a version of $g_3$ in a repair, which is not necessarily a minimum geometry.
For instance, in Figure~\ref{treeRepair} the minimum version of $g_3$ is $g'_{3_1}$. For all other non-minimum versions of $g_3$ in the leaves,
conflicting areas are taken from other geometries. For example, geometry $g'_{3_6}$ results
by keeping $g_3$ as originally and shrinking geometries $g_1$, $g_2$ and $g_4$.

\begin{figure}[!h]
\begin{center}
\begin{tabular}{c}
\includegraphics[width=10cm]{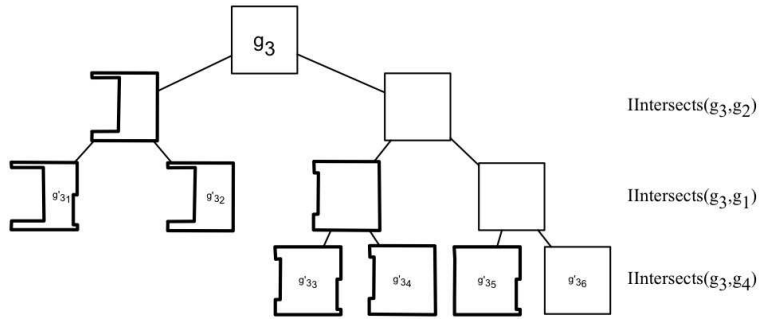}
\end{tabular}
\caption{A tree-based representation of  derived geometries from $g_3$ in some of the  possible minimal repairs (thick boundaries represent geometries that have changed)}
\label{treeRepair}
\end{center}
\end{figure}

The following lemma establishes that when a geometry $g$ is
involved in conflicts of SICs of the form~(\ref{eq:phicore}), there exists a version of $g$ in the repairs that is minimum with respect to set-theoretic (geometric) inclusion. This result is
useful to show that the minimum version of $g$  is the one
that will be in the core.

We need to introduce the set ${\cal G}_{R,\Psi}(\bar{a},g)$
that contains, for a given tuple $R(\bar{a};g)$ in a database
instance $\D$, all the possible versions of geometry $g$ in the
minimal repairs of $\D$.

\begin{definition} \em
Let $\D$ a database instance, a set $\Psi$ of SICs of the
form~(\ref{eq:phicore}) and a fixed tuple $R(\bar{a};g) \in
\D$. Then,  ${\cal G}_{R,\Psi}(\bar{a},g) = \{ g' |
R(\bar{a};g') $ $\in  \D', \D' \in \nit{Rep}(\D,\Psi) \},$
$f^{-1}(R(\bar{a};g')) = R(\bar{a};g)$. \boxtheorem
\end{definition}

\begin{lemma}\em \label{prop:all}
The set of geometries ${\cal G}_{R,\Psi}(\bar{a},g)$ has a minimum element $g_{min}$ under set-theoretic inclusion.
\end{lemma}

\hproof{\noindent By properties of SICs of the form~(\ref{eq:phicore}),
for each conflict in which $R(\bar{a};g) \in \D$ participates,
we can or cannot shrink $g$. This leads to a combination of
possible transformations over geometry $g$ that can be
represented in a binary tree as  shown in
Figure~\ref{treeRepair}.  So, we have a non-empty set of
geometries  ${\cal G}_{R,\Psi}(\bar{a},g)$.

In this tree, we
always have a path from the root to a leaf in which the
geometry is always shrunk; that is, all conflicting areas are
eliminated from $g$. The leaf geometry in this path (repair) is the minimum geometry $g_{min}$. }

\begin{corollary} \label{cor:min}
\em Consider a database instance $\D$, a set $\Psi$ of SICs of the
form~(\ref{eq:phicore}), and a fixed
tuple $R(\bar{a};g) \in \D$. For
$g_{min}$, the minimum geometry in ${\cal
G}_{R,\Psi}(\bar{a},g)$, it holds $R(\bar{a};g_{min}) \in \D^{\star}$.
\end{corollary}

\hproof{\noindent Direct from  Lemma~\ref{prop:all} and  the definition of the core as a geometric intersection.}

\subsection{Properties of the Core}
In this section we establish that for the set of SICs of the
form~(\ref{eq:phicore}), and basic conjunctive queries, it is possible to compute consistent answers on the basis of the core of an inconsistent instance, avoiding the computation of queries in every minimal repair. This is established in Theorems \ref{the:equivalent} and  \ref{the:equivalent2}, respectively.
\begin{theorem} \em \label{the:equivalent}
For an instance $\D$, a set $\Psi$ of SICs of the
form~(\ref{eq:phicore}), and a basic spatial range query ${\cal
Q}(\bar{u};s)$, it holds $\langle \bar{a};g \rangle \in {\nit
Con}({\cal Q},\D,\Psi)$  if and only if $\langle
\bar{a};g\rangle \in {\cal Q}(\D^*)$.
\end{theorem}

\hproof{The projection of range queries always includes the key
of the relation in the result. Thus,  if
$\langle\bar{a},g\rangle \in {\nit Con}({\cal Q},\D,\Psi)$,
then for every $\D' \in \nit{Rep}(\D,\Psi)$, there exists
$R(\bar{b};g')$, such that  $\bar{a} \subseteq \bar{b}$,
$f^{-1}(R(\bar{b};g')) = R(\bar{b};g)$ and $R(\bar{b};g) \in
\D$, where $T(g',w)$ is true for  the spatial constant $w$ of the
range query and $g = \bigcap g'$ with the intersection ranging
over all $g'$.

By Lemma~\ref{prop:all}, there exists tuple $R(\bar{b};g_{min})
\in \D' \in \nit{Rep}(\D,\Psi)$ with $g_{min}  \in {\cal
G}_{R,\Psi}(\bar{b};g)$. If $\langle\bar{a};g\rangle \in
{\nit Con}({\cal Q},\D,\Psi)$, with $\bar{a} \subseteq
\bar{b}$,  $g = \bigcap g' = g_{min}$. Also, it must happen
that  $\langle\bar{a};g_{min} \rangle \in {\cal Q}(\D')$. Then
by  Corollary~\ref{cor:min}, $\langle\bar{a};g_{min} \rangle
\in {\cal Q}(\D^\star)$, and therefore $\langle\bar{a};g
\rangle \in {\cal Q}(\D^\star)$.

In the other direction, if $\langle\bar{a},g^*\rangle \in {\cal
Q}(\D^*)$ (with $\D^* = \bigcap^g \nit{Rep}(\SD,\Psi)$), then
there exists a tuple $R(\bar{b};g^*) \in \D^*$, with $\bar{a}
\subseteq \bar{b}$ and $g^* \neq \go$.  By the monotonicity of
$T \in \{ {\it Intersects, IIntersects}\}$, if $T(g^*,w)$
is true, then for all geometries $g'$  in $R(\bar{b};g') \in
\nit{Rep}(\D,\Psi)$, with $g_{min} \subseteq g'$, $g' \neq
\go$, $T(g',w)$  is also true. Then, by Lemma~\ref{prop:all} and
Corollary~\ref{cor:min}, $g^*= \bigcap g' = g_{min}$ and
$\langle\bar{a},g^*\rangle \in {\nit Con}({\cal Q},\D,\Psi)$.}

\noindent A similar result can be obtained for basic join
queries, i.e., queries that consider two database
predicates (not necessarily different).  Notice that for a SIC $\varphi$ of the form~(\ref{eq:phicore}) with a database predicate $R$ and a basic join query of the form~(\ref{eq:join}) with $R = R_1 = R_2$, the consistent answers do not contain information from tuples that were originally in conflict. This is because by solving conflicts with respect to $\varphi$, all possible intersections between tuples in $R$ will be eliminated (a basic join query asks for geometries that intersect).

The following example illustrates how to
compute consistent answers to basic join queries. This
example will also illustrate the proof of
Theorem~\ref{the:equivalent2}.

\begin{example} \em \label{ex:core4}(example~\ref{ex:core3}  cont.)
Consider the following basic join query posed to the instance $\D$ in Example~\ref{ex:core3}. It is  asking for the identifiers
and geometries of counties and lakes that internally intersect.
\begin{equation}
{\cal Q}(idc,idl;g_1,g_2): \exists n (County(idc,n;g_1)
\wedge Lake(idl;g_2)\wedge {\it IIntersects}(g_1,g_2)). \nonumber
\end{equation}
The consistent answer to this query is $\langle
idc_3,idl_2;g'_{3_1},g'_{7_1}\rangle$. Without using the core,
this answer is obtained by intersecting all  answers
obtained from every possible minimal repair. The geometries in the
repairs of $\D$ with respect to $\Psi$ (SICs~(\ref{eq:phicore1})
and~(\ref{eq:phicore2})) can be partitioned into the following
sets: ${\cal G}_{{\nit County},\Psi}({\nit idc}_1,n_1;g_1) = \{g'_{1_1}, g'_{1_2}\}$, ${\cal G}_{{\nit County},\Psi}({\nit idc}_2,n_2;g_2) = \{\go,g'_{2_1}, g'_{2_2}\}$, ${\cal
G}_{{\nit County},\Psi}({\nit idc}_3,n_3;g_3) = \{g'_{3_1},
g'_{3_2},g'_{3_3},$ $g'_{3_4},g'_{3_5}, g'_{3_6}\}$, ${\cal
G}_{{\nit County},\Psi}({\nit idc}_4,n_4;g_4) = \{g'_{4_1}, g'_{4_2}\}$,  ${\cal G}_{{\nit County},\Psi}({\nit idc}_5,$ $n_5;g_5)$ $= \{g'_{5_1}\}$, ${\cal
G}_{{\nit Lake},\Psi}({\nit idl}_6;g_6) = \{g'_{6_1}, g'_{6_2}\}$,  ${\cal G}_{{\nit Lake},\Psi}({\nit idl}_7;g_7)$ $= \{g'_{7_1}, g'_{7_2}\}$.  The minimum geometries in these seven sets are $g'_{1_1}$, $\go$ (corresponding to the update of  geometry $g_2'$), $g'_{3_1}$, $g'_{4_1}$, $g'_{5_1}$, $g'_{6_1}$, and $g'_{7_1}$, respectively.

Also, for the database predicates ${\nit County}$ and ${\nit
Lake}$, there are two sets containing the possible extensions
of them in the repairs: $\{{\nit County}(\D')| \D' \in {\nit
Rep}(\D, \Psi) \}$, containing the eight versions of counties
(first eight versions of counties in Figure~\ref{coreRepair});
and $\{{\nit Lake}(\D')| \D' \in  {\nit Rep}(\D, \Psi) \}$,
with the two instances of lakes (one with geometries $g'_{6_2}$
and $g'_{7_1}$, and the other with geometries $g'_{6_1}$ and
$g'_{7_2}$ in  Figure~\ref{coreRepair}). Note that the possible minimal repairs
contain combinations of geometries in sets ${\cal G}_{{\nit
County},\Psi}$ $({\nit idc},n;g)$ and ${\cal G}_{{\nit
Lake},\Psi}({\nit idl};g)$.  In particular, there exists a
repair that combines the minimum geometries  $g'_{3_1}$ and
$g'_{6_1}$, and another repair that combines $g'_{3_1}$ and
$g'_{7_1}$.

If the topological  predicate in the basic join query is satisfied by the combination of two minimum geometries, then  other versions of these geometries in other repairs (which geometrically include the minimum geometries) will also satisfy it. In this example, $g'_{3_1}$ and $g'_{7_1}$ intersect and, by the monotonicity  property of predicate {\it IIntersects}, all  other versions of $g_3$ and $g_7$ in other repairs also intersect. As result, $\langle idc_3,idl_2;g'_{3_1},g'_{7_1}\rangle$ is an answer to the query.  Finally,  by Corollary~\ref{cor:min}, $g'_{3_1}$ and $g'_{7_1}$ are in  the core of the database instance and, therefore, $\langle {\nit idc}_3, {\nit idl}_2; g'_{3_1}, g'_{7_1}\rangle$  is also an answer to the query over the core.  \boxtheorem
\end{example}

\begin{theorem} \em \label{the:equivalent2}
For an instance $\D$, a set $\Psi$ of SICs of the
form~(\ref{eq:phicore}), and a basic  spatial join  query
${\cal Q}(\bar{x}_1,\bar{x}_2;s_1,s_2)$, it holds $\langle
\bar{a}_1,\bar{a}_2,g_1,g_2 \rangle \in {\nit Con}({\cal
Q},\D,\Psi)$  if and only if $\langle
\bar{a}_1,\bar{a_2},g_1,g_2\rangle \in {\cal Q}(\D^*)$.
\end{theorem}
\hproof{ The projection of join queries also includes keys.
Thus, if $\langle\bar{a}_1,\bar{a}_2;g_1,g_2\rangle \in {\nit
Con}({\cal Q},\D,\Psi)$, then we have tuples
$R_1(\bar{b}_1;g'_1) \in \D'$, $R_2 (\bar{b}_2;g'_2) \in \D'$,
for every $\D' \in \nit{Rep}(\D,\Psi)$  with $\bar{a}_1
\subseteq \bar{b}_1$, $\bar{a}_2 \subseteq \bar{b}_2$, and
$T(g'_1,g'_2)$ true for $T$ in ${\cal Q}$. Thus, $g_1$ is the intersection of all
those $g'_1$, and $g_2$ is the intersection of all those
$g'_2$.

First, note that if $R_1 = R_2$, only tuples that were not originally in conflict may be in the answer. These tuples will be trivially in the core, because no geometric transformations over their geometries are applied.
Thus, their geometries will be in the answer, if and only if, they satisfy the topological predicate in the query.

By the property (iii) of SICs of the form~(\ref{eq:phicore})
(cf. Remark \ref{re:properties}), solving conflicts on two
different database predicates $R_1$ and $R_2$ are independent.
Let us assume  that $\{R_1(\D')| \D' {\nit Rep}(\D, \Psi) \}$
and $\{R_2(\D')| \D' {\nit Rep}(\D,$ $\Psi) \}$ are the different extensions of predicates $R_1$ and $R_2$ in all possible
minimal repairs. Then, $\nit{Rep}(\D,\Psi)$ contains  database
instances  that result from the combination of these two sets.
Consequently, and using Lemma~\ref{prop:all}, for two given
$\bar{b}_1$ and $\bar{b}_2$, there exists a repair  $\D' \in
\nit{Rep}(\D,\Psi)$ such that $R_1(\bar{b}_1;g'_{1_{min}}) \in
\D'$ and $R_2 (\bar{b}_2;g'_{2_{min}}) \in \D'$, where
$g'_{1_{min}}$ is minimum in ${\cal
G}_{R_1,\Psi}(\bar{b}_1,g_1)$ and $g'_{2_{min}}$ is minimum in
${\cal G}_{R_2,\Psi}(\bar{b_2},g_2)$.

We now prove that if $\langle \bar{a}_1,\bar{a}_2;g_1,g_2
\rangle \in {\nit Con}({\cal Q},\D,\Psi)$, then $\langle
\bar{a}_1,\bar{a_2};g_1,g_2\rangle \in {\cal Q}(\D^*)$.  By
definition of consistent answer, if $\langle
\bar{a}_1,\bar{a}_2;g_1,g_2 \rangle \in {\nit Con}({\cal
Q},\D,\Psi)$, then  $\langle
\bar{a}_1,\bar{a_2},g'_{1_{min}},g'_{2_{min}} \rangle \in {\cal
Q}(\D')$.  By  Corollary~\ref{cor:min},  $\langle
\bar{a}_1,\bar{a}_2;g_1,g_2 \rangle \in {\cal Q}(\D^*)$, with
$g_1 = g'_{1_{min}}$ and $g_2 = g'_{2_{min}}$.

In the other direction, if $\langle
\bar{a}_1,\bar{a}_2,g^*_1,g^*_2 \rangle \in {\cal Q}(\D^*)$,
then $\langle \bar{a}_1,\bar{a}_2,g^*_1,g^*_2 \rangle \in {\nit
Con}({\cal Q},\D,\Psi)$. By Corollary~\ref{cor:min},  $g^*_1 =
g'_{1_{min}}$  and $g^*_2 = g'_{2_{min}}$, and
$R_1(\bar{b}_1;g'_{1_{min}}) \in \D^*$ and $R_2
(\bar{b}_2;g'_{2_{min}}) \in \D^*$. Then, by monotonicity
property of predicate $T \in \{{\it Intersects,}$ ${\it
IIntersects}\}$ in ${\cal Q}$, if  $T(g'_1,g'_2)$ is true, it
is also true for all $R_1(\bar{b}_1;g''_1) \in \D''$ and in
$R_2(\bar{b}_2;g''_2) \in \D''$, with  $\D'' \in
\nit{Rep}(\D,\Psi)$ and $g'_{1_{min}} \subseteq g''_1$ and
$g'_{2_{min}} \subseteq g''_2$.   Therefore, $\langle
\bar{a}_1,\bar{a}_2, g^*_1, g^*_2 \rangle \in {\nit Con}({\cal
Q},\D,\Psi)$. }

\noindent The previous theorems tell us that we can obtain
consistent answer to basic conjunctive queries by direct and
usual query evaluation on the single instance $\D^\star$, the
{\it core} of $\D$.  This does not hold for non-basic
conjunctive queries as the following example shows.

\begin{example} \em \label{ex:nocore}
Consider a database instance with a database predicate $R$
whose geometric attribute values are shown in
Figure~\ref{fig:nocore}(a). This database instance is
inconsistent with respect to a SIC that specifies that
geometries cannot overlap. Let us now consider a range query of
the form $ \exists \bar{y}(R(\bar{x};g) \wedge Touches(g,s))$,
where $s$ is a user defined spatial window, and $\bar{y}
\subseteq \bar{x}$. Figure~\ref{fig:nocore}(b) shows the query
over the intersection of all repairs (the {\it core}),
obtaining geometries $g^{*}_1$ and $g^{*}_2$, from where only $g^{*}_1$ touches $s$. Figures~\ref{fig:nocore}(c) and (d) show
the query over each repair, separately. The answer from the
repair in (c) is $g'_1$, and repair (d) does not return an answer because none of the geometries in this repair touches $s$. Their intersection, therefore, is empty and
differs from the answer obtained from the {\it core}. This
difference is due to the fact that the query window $s$
touches geometry $g'_1$ in only one of the repairs.
\boxtheorem \end{example}

\begin{figure}[t]
\begin{center}
\begin{tabular}{cc}
\includegraphics[width=2.3 cm]{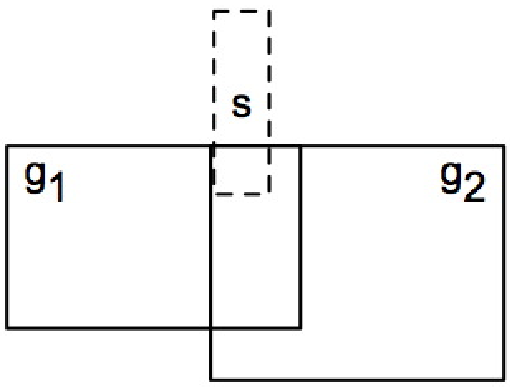}&\includegraphics[width=2.3
cm]{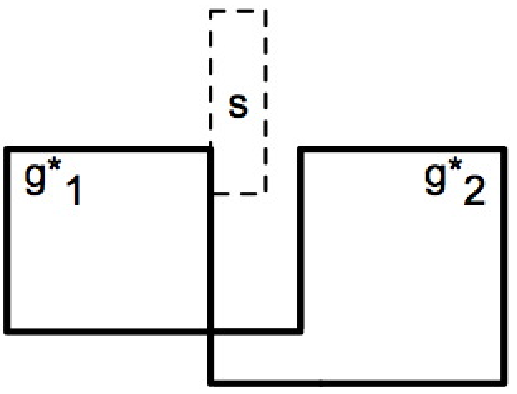}\\
(a)&(b)\\
\includegraphics[width=2.3
cm]{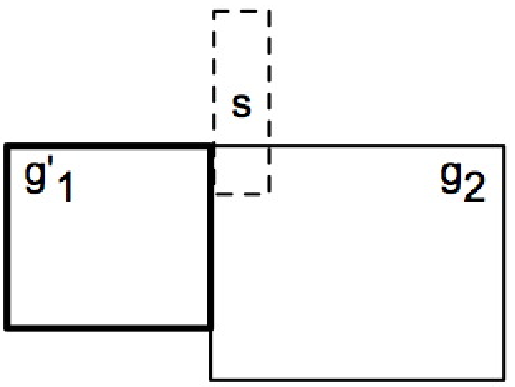}&\includegraphics[width=2.3
cm]{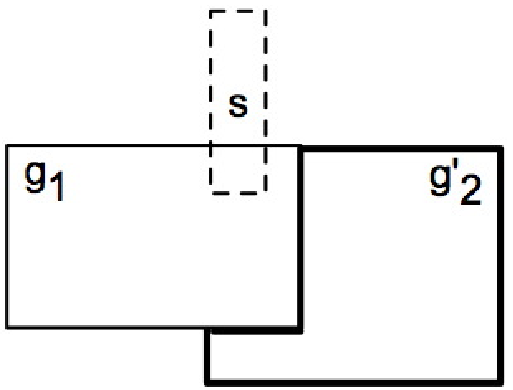}\\
(c)&(d)
\end{tabular}
\caption{~Core vs. consistent answers} \label{fig:nocore}
\end{center}
\end{figure}

\subsection{Computing the Core}\label{sec:computingCore}
We now give a characterization of the core of a database
instance for a set of SICs of the form~(\ref{eq:phicore}), which is not explicitly based on the computation of minimal repairs. This equivalent and alternative characterization of the core allows us to compute the core without having to compute all the minimal repairs.

To simplify the notation, we introduce a logical formula that captures a conflict around a tuple of relation $R \in \SD$ and a SIC of the form~(\ref{eq:phicore}) with topological predicate $T$:
\begin{eqnarray}
\forall \bar{x}_1\bar{x}_2s_1s_2({\it Confl}_{\SD,R,T}(\bar{x}_1,s_1,\bar{x}_2,s_2)
&\equiv& (R(\bar{x}_1;s_1) ~\wedge~ R(\bar{x}_2;s_2) ~\wedge \bar{x}_1 \neq \bar{x}_2~ \wedge~ {\it T}(s_1,s_2))).
\end{eqnarray}

\begin{definition}\em \label{the:core2}
Let $\D$ be a database instance and $\Psi$ a set  of SICs of
the form (\ref{eq:phicore}). For the {\em core} $\D^\star$ of
$\D$ with respect to $\Psi$, it holds  $\D^\star =
\D^\star_{{\it IIntersects}} \cap \D^\star_{{\it Intersects}}
\cap \D^\star_{{\it Equal}}$, where:
\begin{itemize}
\item[(a)] $\D^\star_{{\it IIntersects}} = \{R(\bar{a};{\it Difference}(g,t)) ~|~ R(\bar{a};g) \in
    \D, t = \bigcup \{g' ~|~\textrm{for every} R(\bar{b},g') \in \D \textrm{ such that } \D \models {\nit Confl}_{\SD,R,{\it IIntersects}}(\bar{a},g,\bar{b},g')\}\}$, where
    $\bigcup$ is the {\it geomUnion} operator that calculates the geometric union (spatial aggregation) of geometries.
\item [(b)] $\D^\star_{{\it Intersects}} = \{ R(\bar{a};{\it Difference}(g,t)) ~|~ R(\bar{a};g) \in \D, t = \bigcup \{{\it Buffer}(g',d)$ $~|~ \textrm{for every }$ $ R(\bar{b},g') \in \D \textrm{ such that } \D \models {\nit Confl}_{\SD,R,{\it
Intersects}}(\bar{a},g,\bar{b},g')\}\}$.
\item [(c)] $\D^\star_{{\it Equal}} = \{R(\bar{a};g) ~|~ R(\bar{a},g) \in \D, \textrm{it does not exists } (R(\bar{b},g) \in \D,  \D  \models {\nit Confl}_{\SD,R,{\it
Equal}}(\bar{a},g,\bar{b},g))\}$.\boxtheorem
\end{itemize}
\end{definition}

\noindent Notice that $t$ is the union of all the geometries
that are in conflict with a given geometry $g$. It is obtained
by using the spatial aggregation operator {\it geomUnion}.

Now, we give the specification of the cores: $\D^\star_{{\it
Intersects}}$, $\D^\star_{{\it IIntersects}}$,\footnote{In
current SQL Language ${\nit IIntersects(g_1,g_2) = {\nit
Intersects}(g_1,g_2) \textrm{ AND NOT} {\nit
Touches}(g_1,g_2)}= {\nit Overlaps}(g_1,g_2)$ $ \textrm{ OR }
{\nit Within}(g_1,g_2) \textrm{ OR } {\nit Contains}(g_1,g_2)
\textrm{ OR } {\nit Touches}(g_1,g_2).$} and $\D^\star_{{\it
Equal}}$ as views in a spatial SQL
language.\footnote{Optimizations to the SQL statements are
possible by using materialized views and avoiding double
computation of join operations.} In the following
specification, we assume a database instance $\D$  with a
relational predicate $R({\it id;geometry})$ and primary key $id$.  The following specification shows that our methodologies could be implemented on top of current spatial database management systems.  In particular, the definition of $\D^{\star}_{{\it Intersects}}$ uses a fixed
value $d$ that represents the minimum distance between
geometries in the cartographic scale of the database instance.
The intersection of these views makes $\D^{\star}$.

Table~\ref{views:cores} shows three views that enables to
compute the  core of the database with a
database predicate $R({\it idl;geometry})$.

\begin{table}[h!]
\begin{center}
\begin{scriptsize}
\begin{tabular}{|ll|}
\hline
$\D^\star_{{\it Intersects}}$&
\begin{tabular}{ll}
CREATE VIEW &${\it Core\_Intersects}$\\
AS $($SELECT & ${\it r_1.id}$ AS  ${\it id}$, {\it difference}(${\it r_1.geometry}$, \\
& {\it Buffer}({\it geomunion}(${\it r_2.geometry}),d$)) AS  ${\it geometry}$ \\
FROM  &$R$  AS  $r_1$, $R$ AS  $r_2$\\
WHERE  & ${\it r_1.id <> r_2.id}$   AND \\
&{\it Intersects}(${\it r_1.geometry,r_2.geometry}$) \\
GROUP BY& ${\it r_1.id}$, ${\it r_1.geometry}$\\
UNION\\
SELECT &${\it r_1.id}$  AS  ${\it id}$, ${\it r_1.geometry}$  AS  ${\it geometry}$ \\
FROM&$R$ AS  $r_1$\\
WHERE &NOT EXISTS  (SELECT ${\it r_2.id, r_2.geometry}$ \\
&FROM  $R$ AS  $r_2$ \\
&WHERE ${\it r_1.id <> r_2.id}$   AND\\
&{\it Intersects}$({r_1.geometry,r_2.geometry})))$
\end{tabular}\\
\hline
$\D^\star_{{\it IIntersects}}$&
\begin{tabular}{ll}
CREATE VIEW &${\it Core\_IIntersects}$\\
AS $($SELECT & ${\it r_1.id}$ AS  ${\it id}$, {\it difference}(${\it r_1.geometry}$, \\
&{\it geomunion}(${\it r_2.geometry}))$ AS  ${\it geometry}$ \\
FROM  &$R$  AS  $r_1$, $R$ AS  $r_2$\\
WHERE  &  ${\it r_1.id <> r_2.id}$   AND \\
&{\it Intersects}(${\it r_1.geometry,r_2.geometry}$) AND \\
&NOT {\it Touches}(${\it r_1.geometry,r_2.geometry}$) \\
GROUP BY& ${\it r_1.id}$, ${\it r_1.geometry}$ \\
UNION\\
SELECT &${\it r_1.id}$  AS  ${\it id}$, ${\it r_1.geometry}$  AS  ${\it geometry}$ \\
FROM&$R$ AS  $r_1$\\
WHERE &NOT EXISTS  (SELECT ${\it r_2.id,r_2.geometry}$ \\
&FROM  $R$ AS  $r_2$ \\
&WHERE   ${\it r_1.id <> r_2.id}$   AND\\
&{\it Intersects}(${\it r_1.geometry,r_2.geometry}$) AND\\
&NOT {\it Touches}(${\it r_1.geometry,r_2.geometry})))$
\end{tabular}\\
\hline
$\D^\star_{{\it Equal}}$&
\begin{tabular}{ll}
CREATE VIEW& ${\it Core\_Equal}$ \\
AS $($SELECT & ${\it r_1.id}$ AS  ${\it id}$, ${\it r_1.geometry}$  AS  ${\it geometry}$ \\
FROM&$R$  AS  $r_1$\\
WHERE& NOT EXISTS (SELECT  ${\it r_2.id,r_2.geometry}$ \\
&FROM  $R$  AS  $r_2$ \\
&WHERE  ${\it r_1.id <> r_2.id}$  AND  \\
&{\it Equals}$({\it r_1.geometry,r_2.geometry})))$
\end{tabular}\\
\hline
\end{tabular}
\end{scriptsize}
\caption{SQL statements to compute views for  $\D^\star_{{\it
Intersects}}$,  $\D^\star_{{\it IIntersects}}$, and
$\D^\star_{{\it Equal}}$} \label{views:cores}
\end{center}
\end{table}

\begin{example} \em (example \ref{ex:CQA2} cont.) \label{ex:coreCQA}
The example considers only the relation {\it LandP} with primary key $idl$ and the SIC (\ref{eq:constraint}) of Example~\ref{ex:SDB}.  We want to consistently answer the query of Example~\ref{ex:core}, i.e., $\exists \nit{name~ owner}
(\nit{LandP}$ ${\nit (idl,name,owner;geometry)}~ $ $\wedge$ ${\nit IIntersects(geometry},([x_1,y_1],[x_2,y_1],[x_2,y_2],[x_1,y_2],[x_1,y_1])).$

To answer this query, we generate a view of the ${\it core}$
applying the definition in Table~\ref{views:cores}. That is,
we eliminate from each geometry the union of conflicting
regions with respect to each land parcel. In this case, the
conflicting geometries for $g_2$ are $g_3$ and $g_4$; for
geometry $g_3$ is $g_2$; and for geometry $g_4$ is $g_2$. This
is the definition of the core in SQL:

\begin{scriptsize}
\begin{eqnarray}
&\textrm{CREATE VIEW  }& Core \nonumber \\
&\textrm{AS (SELECT }& l_1.idl \textrm{ AS } idl, l1.name \textrm{ AS } name, l_1.owner \textrm{ AS } owner, \nonumber \\
&&{\it difference}(l_1.geometry, {\it geomunion}(l_2.geometry)) \textrm{ AS } geometry \nonumber\\
&\textrm{FROM  }&LandP \textrm{ AS } l_1, LandP \textrm{ AS } l_2 \nonumber \\
&\textrm{WHERE  }&l_1.idl <> l_2.idl  \textrm{ AND } {\it Intersects}(l_1.geometry,l_2.geometry) \textrm{ AND} \nonumber\\
&&\textrm{NOT} {\it Touches}(l_1.geometry,l_2.geometry) \nonumber\\
&\textrm{GROUP BY}& l_1.idl,l_1.name,l_1.owner, l1.geometry \nonumber\\
&\textrm{UNION }&\nonumber \\
&\textrm{SELECT }&l_1.idl  \textrm{ AS } idl, l_1.name \textrm{ AS } name, l_1.owner \textrm{ AS } owner, l_1.geometry \textrm{ AS } geometry \nonumber \\
&\textrm{FROM  }&LandP \textrm{ AS } l_1 \nonumber \\
&\textrm{WHERE}  &\textrm{NOT EXISTS} \textrm{(SELECT } l_2.idl,l_2.geometry \nonumber \\
&&\textrm{FROM  } LandP \textrm{ AS } l_2 \nonumber \\
&&\textrm{WHERE } l_1.idl <> l_2.idl  \textrm{ AND } {\it Intersects}(l_1.geometry,l_2.geometry) \textrm{ AND} \nonumber\\
&&\textrm{NOT} {\it Touches}(l_1.geometry,l_2.geometry))) \nonumber
\label{view:CLandparcel}
\end{eqnarray}
\end{scriptsize}
We now can pose the query to  the core to compute the consistent answer to the original query:

\begin{eqnarray}
&\textrm{SELECT }& {\it idl,name,owner,geometry}\\&\textrm{FROM  }& {\it Core} \nonumber \\
&\textrm{WHERE  }&{\it Intersects}({\it geometry},([x_1,y_1],[x_2,y_1],[x_2,y_2],[x_1,y_2],[x_1,y_1]))\nonumber
\label{query:range}  \vspace{-8mm}
\end{eqnarray}

The answer is shown in Figure~\ref{fig:CQALand}. This query is a classic selection from the ${\it Core}$ view. \boxtheorem
\end{example}

\noindent This core-based method allows us to compute consistent
answers in polynomial (quadratic) time (in data
complexity) in cases where there can
be exponentially many repairs. In Example~\ref{ex:PTA3}, where we
have $2^{n-1}$  minimal repairs, we can apply the query ${\cal Q}$
over the {\it core}, and we only have to compute the difference of a
geometry with respect to the union of all other geometries in
conflict. This corresponds to a polynomial time algorithm of order
polynomial with respect to the size of the database instance.

\section{Experimental Evaluation}\label{sec:experiments}
In this section we analyze the results of the experimental
evaluation we have done of the core-based CQA using synthetic
and real data sets. The experiment includes a scalability
analysis that compares the cost of CQA with increasing numbers
of conflicting tuples and increasing sizes of  database
instances. We compare these results with respect to the direct
evaluation of basic conjunctive queries over the inconsistent
database (i.e., ignoring inconsistencies). The latter reflects
the additional cost of computing consistent answers against
computing queries that ignore inconsistencies.

\subsection{Experimental Setup}
We create synthetic databases to control the size of the
database  instance and the number of conflicting tuples. We use
a database schema consisting of a single predicate $R({\nit
id};{\nit geometry})$, where $\nit{id}$ is the numeric key and
$\nit{geometry}$ is a spatial attribute of type polygon. We
create three sets of synthetic database instances by
considering SICs of the form~(\ref{eq:phicore}) with different
topological predicates:

\begin{center}
\begin{tabular}{|ll|}\hline
Set& \multicolumn{1}{c|}{SIC}\\ \hline \hline
Equals&$\overline{\forall}~\neg(R(\bar{x_1};s_1) ~\wedge~
R(\bar{x_2};s_2) ~\wedge~ \bar{x_1} \neq \bar{x_2}~
 \wedge~\it{Equals}(s_1,s_2))$\\
Intersects&$\overline{\forall}~\neg(R(\bar{x_1};s_1) ~\wedge~
R(\bar{x_2};s_2) ~\wedge~ \bar{x_1} \neq \bar{x_2}~
 \wedge~\it{Intersects}(s_1,s_2))$\\
IIntersects&$\overline{\forall}~\neg(R(\bar{x_1};s_1) ~\wedge~
R(\bar{x_2};s_2) ~\wedge~ \bar{x_1} \neq \bar{x_2}~
 \wedge~\it{IIntersects}(s_1,s_2))$\\
 \hline
\end{tabular}\end{center}

\vspace{0.3 cm}

\noindent For each set we create five consistent instances
including 5,000, 10,000, 20,000, 30,000, and 40,000 tuples of
homogeneously distributed spatial objects whose geometries are
rectangles (i.e., 5 points per geometric representation of
rectangles). Then, we create inconsistent instances with
respect to the corresponding SICs in each set with 5\%, 10\%,
20\%, 30\%, and 40\% of tuples in conflict. For database
instances with a SIC and topological ${\it Equals}$, we create
inconsistencies by duplicating geometries in a percentage of
geometries. For database instances with a SIC and topological
${\it IIntersects}$, we create inconsistencies by making geometries overlap. Finally, for database
instances with a SIC and topological ${\it Intersects}$, we
create inconsistencies by making a percentage of geometries to
touch.

Due to the spatial distribution of rectangles in the
sets, the  cores for  database instances with
SICs using topological predicates in $\{{\it Intersects,
IIntersects}\}$ have the same numbers of   points  in their
geometric representations  than  their original  instances. For the
set of database instances with  SICs using topological
predicate {\it Equals}, the numbers of points in the geometric
representations of their cores are less than in the
original databases, because we eliminate geometries as we
restore consistency. Thus, we are not introducing additional
storage costs in our experiments.

To have a better understanding of the computational cost of
CQA, we also evaluate the cost of CQA over real and free
available data of administrative boundaries of Chile
\cite{datachile}. Chilean administrative boundaries have
complex shapes with many islands, specially, in the South of
Chile (e.g.,  a region can have  891 islands).   For the real
database, we have two predicates ${\it Counties}$ and  ${\it
Provinces}$. Notice that, at the conceptual label,  ${\it
Provinces}$ are aggregations of ${\it Counties}$. In this
experiment, however, we have used the source data as it is,
creating  separated tables for ${\it Counties}$ and  ${\it
Provinces}$ with independent spatial attributes.  For this real database, we consider  SIC of
the form: $\overline{\forall}~\neg({\it R}(\bar{x_1};s_1)
~\wedge~{\it R}(\bar{x_2};s_2) ~\wedge~ \bar{x_1} \neq
\bar{x_2}~\wedge~\it{IIntersects}(s_1,s_2))$, with $R$ being
${\it Counties}$ or  ${\it Provinces}$.

Table~\ref{tab:data} summaries the data sets for the
experimental evaluation. The percentage of inconsistency is
calculated as the number of tuple in any conflict over the
total number of tuples. The geometric representation size is
calculated as the number of points in the boundaries of a
region.

\begin{table}[ht]
\centering \begin{tabular}{|l|llll|} \hline
Source&Name&Tuples&Inconsistency (\%)&Geometric representation size\\
\hline\hline
Synthetic&Equals&5,000-40,000&5-40&25,000-200,000\\
\cline{2-5}
&IIntersects&5,000-40,000&5-40&25,000-200,000\\
\cline{2-5}
&Intersects&5,000-40,000&5-40&25,000-200,000\\
\hline
Real&Provinces&52&59&35,436\\
\cline{2-5}
&Counties&307&12.7&72,009\\
\hline
\end{tabular}
\caption{Data sets of the experimental evaluation} \label{tab:data}
\end{table}

We measure the computational cost in terms of seconds needed to
compute the SQL statement on a Quad Core Xeon X3220 of 2.4 GHz,
1066 MHz, and 4 GB in RAM. We use as spatial DBMS PostgreSQL 8.3.5
with PostGIS 1.3.5.

\subsection{Experimental Results}

Figure~\ref{fig:graficoCore} shows the cost of the core
computation for the different synthetic database instances. To
make this experimental evaluation easier and faster,  we used
materialized views  so that we computed only once the core and
applied queries on this core's view. However, we added the
computational cost of the core to each individual query result
to have a better understanding of the  cost of applying CQA.

The time cost of
computing the core for inconsistent databases with respect to a
SIC with a topological predicate ${\it Equals}$ decreases as
the number of tuples in conflicts increases, since the core
eliminates geometries in conflict and, therefore, these empty
geometries are then  ignored  in geometric computations.  The
cost of computing the core is largely due to the spatial join given by the topological predicate of a SIC,
which could decrease using more efficient algorithms and
spatial indexing structures.

\begin{figure}[h!]
\begin{center}
\includegraphics[width=9 cm]{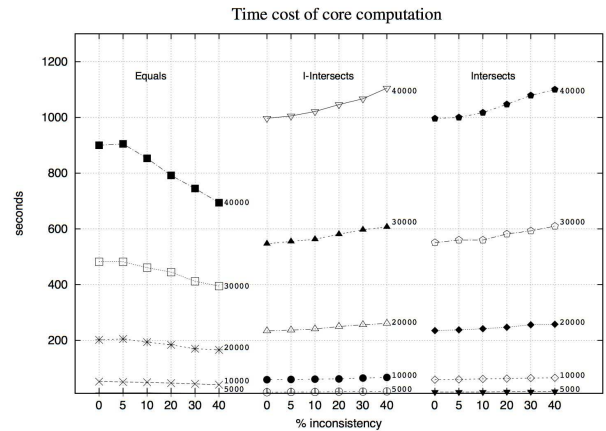}
\caption{Time cost of the core computation for different SICs,
different levels of inconsistency, and different sizes of databases instances}
\label{fig:graficoCore}
\end{center}
\end{figure}

For the synthetic database instance, Figures~\ref{fig:graficoRange} and~\ref{fig:graficoJoin} show the cost rate between computing a CQA with respect to simple range or join queries (with the spatial predicate {\it
Intersects}) that ignore inconsistencies. Range queries use a
random query window created by a rectangle whose side is equivalent to 1\% of the total length in each dimension. Notice that the time cost of computing a range query for a database instance with 10,000 was approximately 15 ms, which, in average, was 900 times less than computing a join query. These reference values exhibit linear and quadratic growth  for range and join queries, respectively, as we consider increasing
sizes of database instances.  The computational cost of CQA to join queries include the computation of the core; however, this  cost could
be amortized if we use a materialized view of the core
 for computing more than one join query. In the time cost of CQA for
range queries, we have optimized the computation by applying
the core-computation over a subset of tuples previously
selected by the query range. This optimization is not possible
for join queries, since no spatial window can constrain the
possible geometries in the answer.

\begin{figure}[h!]
\begin{center}
\includegraphics[width=9 cm]{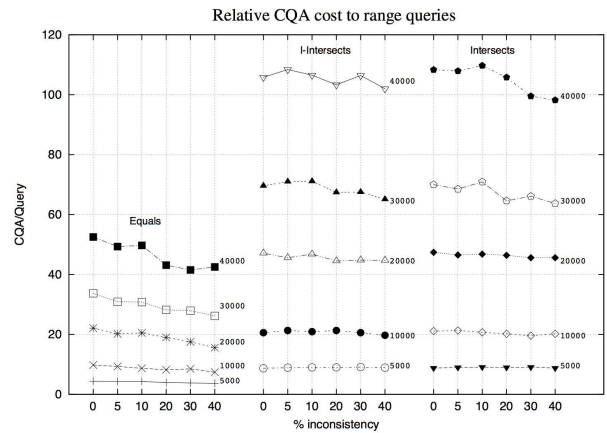}
\caption{Relative cost of CQA to range queries}
\label{fig:graficoRange}
\end{center}
\end{figure}

\begin{figure}[h!]
\begin{center}
\includegraphics[width=9 cm]{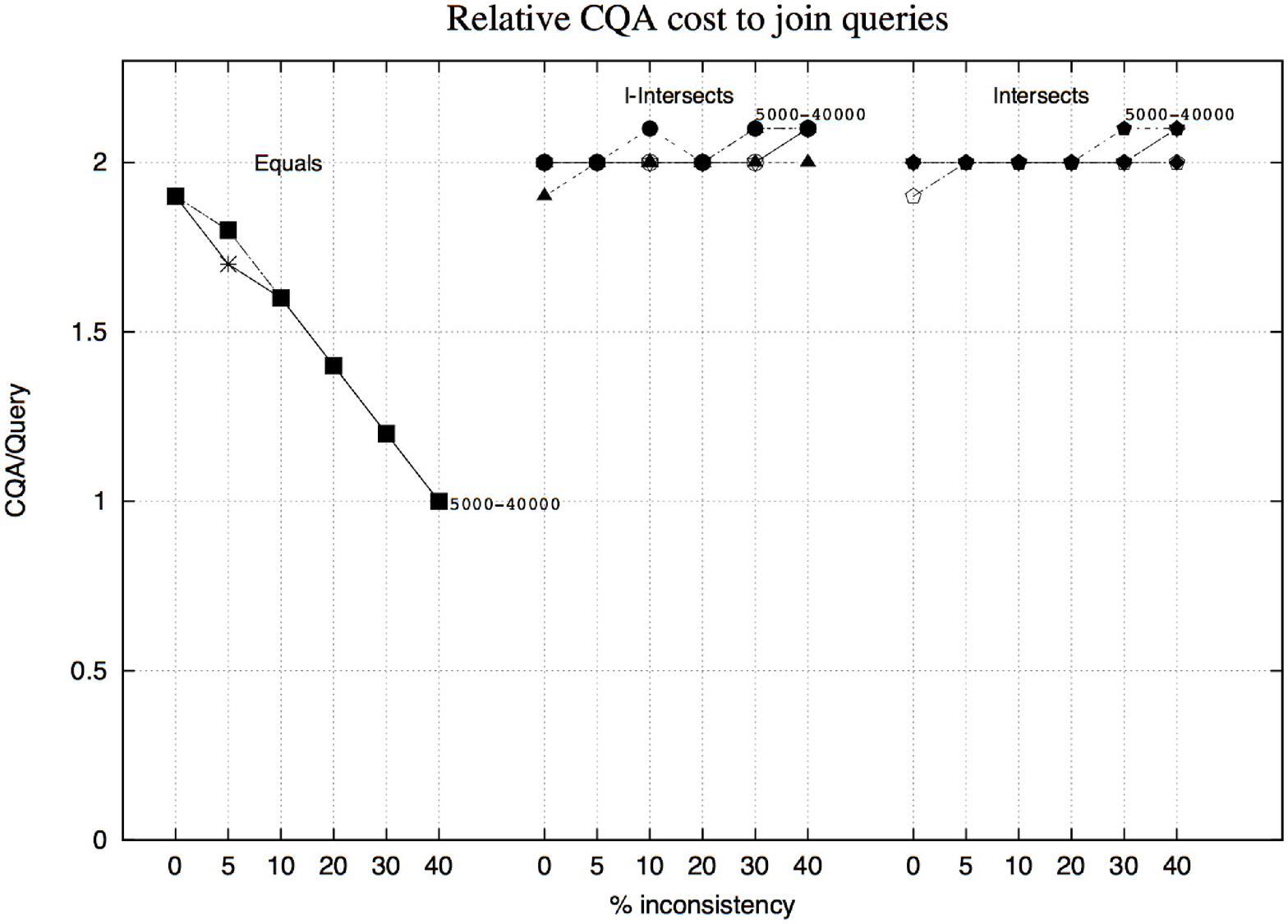}
\caption{Relative cost of CQA to join queries}
\label{fig:graficoJoin}
\end{center}
\end{figure}

The results indicate that CQA to a range query can cost 100 times the cost
of a simple query. This is primarily due to the join
computation of the core. Indeed, when comparing the CQA to a
join query, we only duplicate the relative cost, and in the
best case, keep the same cost. However, join queries have a
significant larger computational cost.   Notice that the
computation cost for a CQA to range query is around 60s in the
worst case (40,000 tuples). With exception of cases when the
core contains empty geometries, the percentage of
inconsistencies does not affect drastically the results.

We also evaluate the scalability of the  CQA cost to range
queries in function of the size of the query window (i.e.,
spatial window). In Figure ~\ref{graficoRange2} we show the
relative CQA cost to range queries on a synthetic database
instance with 10,000 tuples and  range queries whose random
spatial windows varied from 1\% to 5\% of the size in each
dimension. The results indicate that the relative cost
increases logarithmical  as we increase the size of the query
window. Also, only for database instances with a SIC and
topological predicate ${\it Equals}$, the relative cost suffers
some variation across different percentages of inconsistencies,
primarily, due to the elimination of geometries in the
database.

\begin{figure}[h!]
\begin{center}
\includegraphics[width=9 cm]{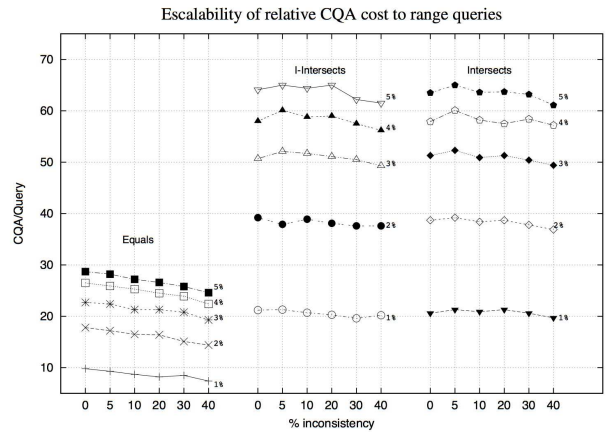}
\caption{Relative cost of CQA to range queries and different sizes of the query window
(using a database instance with 10,000 tuples )}\label{graficoRange2}
\end{center}
\end{figure}

Finally,  we applied the core-based computation of CQA to the
real database instances in Table~\ref{tab:data}.
Table~\ref{tab:real} summaries the results obtained with these
data, which were in agreement with the results obtained with
the synthetic database instances.  In this table, $\Delta {\it
Points}$ represents the relative difference in the size of the
geometric representation between the core and the original
database.  Notice that computing the core increased the
geometric representation of ${\it Provinces}$ up to 5.0\%,
which is bounded by the shape of geometries  in conflict (i.e.,
the size of the  original geometric representation).   In
the case of ${\it Counties}$, however, the size of the
geometric representation of the core decreases down to
$-0.03\%$. Since  the geometry of provinces should be the
geometric aggregation of counties, we could expect to have a
relationship between $\Delta {\it Points}$ for ${\it
Provinces}$  and ${\it Counties}$. However, the source data set
uses independent geometries for ${\it Provinces}$  and ${\it
Counties}$ and no comparison can be made.

\begin{table}[ht]
\centering \begin{tabular}{|l|llllll|} \hline
&&&\multicolumn{2}{c}{{  Range}}&\multicolumn{2}{c|}{{ Join}}\\
{ Data}&{$\Delta$ Points}&{  Core}&{  Simple}&{  CQA}&{  Simple}&{  CQA}\\
\hline\hline
Provinces&+5.0\%&17.7&0.04&0.25&29.8&63.4\\
Counties&-0.03\%0&18.1&0.1&2.1&40.6&55.7\\
\hline
\end{tabular}
\caption{CQA cost with real data (costs of core and queries in
seconds)} \label{tab:real}
\end{table}

\section{Conclusions}\label{se:conclusions}

We have formalized a repair semantics and consistency query
answers for spatial databases with respect to SICs. The repair
semantics is used as an auxiliary concept for handling
inconsistency tolerance and computing consistent answers to
spatial queries. It is based on updates that shrink geometries
of objects, even at the point of deleting geometries for some
exceptional cases, as for predicate {\it Disjoint}. Geometries
are virtually updated applying admissible geometric operators,
which are available in most spatial DBMSs.

By restricting ourselves to the application of the admissible
transformations, we have a finite number of possibilities for
making a pair of geometries consistent with respect to a SIC.
However, there may still be exponentially many repairs for a
given instance and set of SICs. With the purpose of avoiding to
compute and query all repairs, we have identified cases of SICs
and conjunctive (range and join)  queries where the consistent
answers can be obtained by posing a standard query to a single
view of the original instance.  This view is equivalent  to the
intersection of all possible minimal repairs, what we called the ${\it
core}$ of a database instance,  which for a subset of SICs can
be computed in polynomial time without determining each repair.

An experimental evaluation of the core-based computation of CQA
reveals that answering range queries has a cost that varies
drastically in function  of the topological predicates in SICs
and the number of tuples in the database instance, reaching up
to 100 times the cost of a simple range query. This is mainly
due to the spatial join involved in computing the core.  For
join queries, instead, the cost of CQA is the double of a
simple join query.  These results do not use optimizations with
spatial indexing, which has been left for future work. Even
more, they assume that we have to compute the core for each
query, which could be optimized by using materialized
views.

This work leaves many problems open.  Most prominently,
computability and complexity issues have to be explored. For
example, some interesting decision problems are  deciding if
non trivial repairs (i.e., not obtained by cancellation of
geometries) exist for an instance and a set of SICs, or
deciding whether or not a particular instance is a repair of an
inconsistent database instance. The complexity of deciding if a
spatio-relational tuple is a consistent answer is also open. As in the relational case, we
expect to find hard cases for all these problems. For them, it
would be interesting to obtain lower complexity approximation
algorithms.

We have considered only regions to represent spatial objects. A
natural extension of this work would be to define a repair
semantics for other spatial abstractions, such as polylines,
points, networks, and so on. We would also like to explore not
only denial SICs, but also other classes of semantic ICs, and other types of repair semantics that include solving conflicts with respect to a topological predicate ${\nit Disjoint}$. This
includes also the possibility of considering combinations of
spatial with relational constraints, e.g. functional
dependencies and  referential ICs.

\subsection*{Acknowledgments}

This project is partially funded by  FONDECYT, Chile, grant
number 1080138. Part of this research was done when L. Bertossi
was invited to the Universidad de Concepcion and Universidad
del B\'io-B\'io. M\'onica Caniup\'an received  funding from
FONDECYT, Chile, grant number 11070186. Leopoldo Bertossi has
also been partially funded by an NSERC Discovery Grant
(\#315682).
\bibliographystyle{acm}

\end{document}